\documentclass[twocolumn,showpacs,preprintnumbers,aps,prc,10pt,nofootinbib]{revtex4-1}

\usepackage[english]{babel} 	
\usepackage{graphics} 
\usepackage{graphicx} 			
\usepackage{amssymb} 			
\usepackage{amsthm} 			
\usepackage{amsmath} 			
\usepackage{siunitx} 			
\usepackage{subfig} 			

\usepackage{dcolumn}
\newcolumntype{d}[1]{D{.}{.}{#1} }

\begin{document}

% ----------------------     Authors   ------------------------------ %

\title{Borromean structures in medium-heavy nuclei}

\author{D. Hove, D.V. Fedorov, H.O.U. Fynbo, A.S. Jensen, K. Riisager, 
N.T. Zinner}
\affiliation{Department of Physics and Astronomy, Aarhus University, DK-8000 Aarhus C, Denmark} 

\author{E. Garrido}
\affiliation{Instituto de Estructura de la Materia, IEM-CSIC,
Serrano 123, E-28006 Madrid, Spain}

\date{\today}

% ----------------------     Abstract    ------------------------------ %

\begin{abstract}

Borromean nuclear cluster structures are expected at the corresponding
driplines.  We locate the regions in the nuclear chart with the most
promising constituents, it being protons and $\alpha$-particles and
investigate in details the properties of the possible Borromean
two-$\alpha$ systems in medium heavy nuclei.  We find in all cases that
the $\alpha$-particles are located at the surface of the core-nucleus as
dictated by Coulomb and centrifugal barriers.  The two lowest
three-body bound states resemble a slightly contracted
$^{8}\text{Be}$ nucleus outside the core.  The next two excited
states have more complex structures but with strong components of
linear configurations with the core in the middle.  $\alpha$-removal
cross sections would be enhanced with specific signatures for these
two different types of structures.  The even-even Borromean two-$\alpha$
nucleus, $^{142}$Ba, is specifically investigated and predicted to
have $^{134}\text{Te}-\alpha-\alpha$ structure in its ground state and low-lying spectrum.
\end{abstract}

\pacs{21.45.+v, 31.15.xj, 21.60.Gx}

\maketitle

% ----------------------     Main text      ------------------------------ %

\section{Introduction}

Surprisingly large nuclear reaction cross sections were found
experimentally in 1985 \cite{tan85a,tan85b} for light nuclei colliding
with $^{11}$Li.  Normal cross section were found for all lighter
lithium isotopes.  The interpretation and qualitative understanding
were almost immediately explained as based on weak binding in relative
$s$-waves \cite{han87}.  The structure is in general called a nuclear
halo, since the spatial extension is much larger than radii of
ordinary nuclei \cite{riis13}.  The essence of theoretical descriptions is contained
in a schematic model for two short-range interacting point-like
particles as reviewed in \cite{jen04}.  Thresholds for binding nuclear
clusters enhance the probability for finding decoupled and spatially
extended nuclear structures.  Driplines related to different nucleons
or bound clusters of nucleons provide the best environment for the
corresponding ground state structures. The features of such nucleon
dripline nuclei are reviewed in \cite{tho04}.  Nucleons can be
arranged in bound clusters and thereby form the constituents for novel
nuclear few-body structures as reviewed in \cite{oer06,fre07,oko12}.

Increasing the number of clusters progressively reduces the
possibility of large spatial extension when all particles are
distinctly separated.  An effective centrifugal barrier confines the
mean square radius for a bound system, and overlapping nuclei of
finite size would couple intrinsic and relative degrees of freedom.
To prevent this collapse into one much larger many-body system, the
particles may correlate strongly into clusters and effectively reduce
the active degrees of freedom corresponding to much fewer
well-separated clusters of particles \cite{jen04}. Already four
particles with infinitesimally small binding energy have finite
root-mean-square radius, in sharp contrast to diverging radii of two-
and three-body systems approaching zero binding energy \cite{yam11}.

The repulsive centrifugal barrier combined with a short-range
attraction of given radius may leave an attractive pocket able to hold
a bound state.  This prevents occurrence of nuclear halos of large
relative angular momenta.  The same mechanism opposes halos where the
Coulomb repulsion dominates or contributes significantly.  Thus,
nuclear halos are most likely to occur for very small binding energy,
two- and three-body systems, relative $s$ and $p$-waves, and small
pairwise charge products \cite{fed94a,fed94b,jen03}.  Two-body halos
should then be searched for at their threshold for binding while
still subject to these conditions, both for ground and excited states.

Three-body cluster states are probably less frequent and potentially
less pronounced than two-body halos.  However, investigations of
occurrence and properties are essentially all confined to light
nuclei.  For ground states the most promising structures appear to be
for Borromean systems, since the unbound two-body subsystems are
prohibited from reducing the active degrees of freedom to an effective
two-body system.  Then the conditions are positive pairwise cluster
binding and very small three-body binding.  The most obvious
constituents are neutrons, protons, and $\alpha$-particles.  Heavier
particles necessarily have both larger radii and charges, and
therefore less obvious constituents in a halo system.

Pairs of identical nucleons and $\alpha$-particles are always unbound,
and the Borromean properties are therefore determined by the pairwise
binding energies to the third particle \cite{zhu93}. The positive binding of
nucleon-core and a core-$\alpha$ systems dictate the position in the
nuclear chart to be around the corresponding driplines.  As we shall
discuss later, combining nucleons and $\alpha$-particles is then not possible
for neutrons while the proton dripline is suitable for nuclei with
neutron number $N > 30$.  We shall not deal with the individual light
nuclei where the Borromean properties are thoroughly discussed in the
available literature \cite{riis13, fre07}. 

Two $\alpha$-particles and a heavier core along the $\alpha$-dripline can
form a Borromean system.  We shall in the present paper concentrate on
two $\alpha$-particles plus a medium heavy core-nucleus which is a system
so far very little discussed.  The Borromean property is established
experimentally for a few nuclei \cite{aud12} as pointed out recently
\cite{baa13}.  It remains to be seen whether $\alpha$-particles in the
end turn out to be substantially contributing constituents to the
structure of some low-energy states in medium heavy nuclei.  The
minimum requirement is that the intrinsic $\alpha$-particle degrees of
freedom effectively decouple from all other nuclear degrees of
freedom.  Recent theoretical investigations of large systems confirm
the expectation that $\alpha$-particles are advantageous for nuclear
matter densities corresponding to the tail of a nucleus
\cite{ldm97,ebr14}.

Explicit use of $\alpha$-particle degrees-of-freedom is complementary to
mean-field approximations, where correlations only appear through
shell structures of single nucleons (or pairs) in deformed average
fields. Such models provide completely different basis states, but
they are not necessarily unable to describe the same features of some
many-body states. In this context it is interesting to note that
octupole deformation has been an important ingredient in descriptions
of nuclei located close to the $\alpha$-dripline \cite{but96}, where $\alpha$-clustering
is most likely to occur in ground states. We assume $\alpha$-particles
as the basic constituents with the inherent limits of validity that
only very specific structures can be described. However, it may very
well be states that cannot be described in mean-field or shell-models,
or at most only with severe difficulties indicating that an
inappropriate basis is chosen.

The $\alpha$-clusterization, or more moderately $\alpha$-correlations,
should produce an otherwise larger binding energy which then simply
could be measured as the revealing observable.  The increased binding
has to be compared to the surrounding nuclei, and the signal would be
contained in appropriate mass differences \cite{jen84,hov13}.  This
would be in complete analogy to the odd-even staggering related to the
``pairing gaps'' \cite{hov14}.  Unfortunately, such a signal in the
variation of the binding energy between neighbouring nuclei is extremely
difficult to separate from the smooth background variation which
therefore necessarily has to be eliminated.  The optimistic point of
view would be that $\alpha$-correlations are more extended and vary
smoothly over smaller or larger regions of the nuclear chart.

Instead of futile searching for signals in the binding energies
\cite{jen84}, we shall therefore directly calculate three-body
properties from an assumption of the presence of two $\alpha$-particles
surrounding a heavier core-nucleus.  We are then able to study the
solutions, test compatibility with the assumptions, and predict the
properties of the emerging structures.  We first in Sec.~\ref{sec drip} outline
which regions of the nuclear chart are most promising. In Sec.~\ref{sec 2p}
we briefly sketch the computational procedure and specify the necessary
parameters. This also includes the two-body $\alpha$-core potential used in the initial three-body calculations. In Sec.~\ref{sec nd148} $^{148}\text{Nd}$ $(^{140}\text{Ba}+\alpha+\alpha)$ is used as a trial system to evaluate the general nature of such three-body $\alpha$-Borromean systems. Dedicated calculations for the Borromean $^{142}$Ba $(^{134}\text{Te}+\alpha+\alpha)$ nucleus are reported in Sec.~\ref{sec res}, where comparisons are made to experimental results for $\alpha$-dripline nuclei. Finally Sec.~\ref{sec con} contains a summary and the conclusions.

\section{Driplines and Borromean regions \label{sec drip}}

Borromean systems are, almost by definition, most often weakly bound,
since all three two-body subsystems must be unbound and the same
interactions are responsible for the three-body binding.  This
definition is appropriate for ground states of systems where the
cluster division already is made. The total system may very well have
much deeper-lying bound states of different structures where the
cluster division is completely inadequate.  Thus the structures of
interest here can appear as relatively highly excited states of the
given nucleus. 

It is only within the decided cluster structures that the
corresponding three-body system is relatively weakly bound compared to
the threshold of large-distance separation of all three clusters.  The
weak cluster-binding is compatible with large size and with cluster
identities maintained.  Therefore, the most likely region for finding
three-body cluster states should be where the effective
cluster-cluster interaction provides small, positive or negative,
binding energies.  Cluster driplines are then useful in outlining
regions where corresponding Borromean systems should be more likely.

Let us now focus on two $\alpha$-particles surrounding a core-nucleus.
We want to find the $\alpha$-dripline with vanishing $\alpha$-separation
energy, $S_{\alpha}$, that is
\begin{align}
S_{\alpha}(A,X) 
&= B(A+4,X) - B(A,X) - B_{\alpha}, \label{eq a sep}
\end{align}
where $B$ is the nuclear binding energy, $N$ and $Z$ are neutron and
proton numbers, $X = N-Z$, $A = N+Z$, and $B_{\alpha} = 28.295 \,
\si{\mega\electronvolt}$ is the binding energy of an $\alpha$
particle.  The dripline defined by $S_{\alpha}(A,X)=0$ can be
estimated by use of the liquid drop model, or specifically
\begin{align}
 & B_{LD} 
= a_v A - a_s A^{2/3} - a_c \frac{Z^{2}}{A^{1/3}} - a_a \frac{(A-2 Z)^{2}}{A}  \label{eq bind} \\ \notag 
& = a_v A - a_s A^{2/3} - \frac{a_c}{4 A^{1/3}} \left( A^{2} + X^{2} - 2AX \right)  - a_a \frac{X^{2}}{A}, 
\end{align}
where we use the parameter values \cite{ldm97}: $(a_v,a_s,a_c,a_a)=
(15.56, 17.23, 0.7, 23.285)$, all in MeV. Then $S_{\alpha}(A,X)=0$
combined with Eqs.~(\ref{eq a sep}) and (\ref{eq bind}) results in the
quadratic equation
\begin{eqnarray}
0 &=& c_0 + c_1 X + c_2 X^2 \;, \label{drip0} \\ \label{drip1} 
c_0 &=&  4 a_v - B_{\alpha}  \\ \nonumber
 &-& a_s \left( (A+4)^{\frac{2}{3}} - A^{\frac{2}{3}} \right)
- \frac{a_c}{4} \left( (A+4)^{\frac{5}{3}} - A^{\frac{5}{3}} \right)\;, \\ 
\label{drip2}
c_1 &=& \frac{a_c}{2} \left((A+4)^{\frac{2}{3}} - A^{\frac{2}{3}} \right)\;, \\
c_2 &=& -\frac{a_c}{4} \left(\frac{1}{(A+4)^{\frac{1}{3}}} -
\frac{1}{A^{\frac{1}{3}}} \right) 
 - a_a \left( \frac{1}{A+4} - \frac{1}{A} \right) . \label{drip3}
\end{eqnarray}
The two solutions, $X_{\pm}(A)$, to Eq.~(\ref{drip0}) are combined
with $Z = (A- X_{\pm}(A))/2$ and $N= (A + X_{\pm}(A))/2$ to give the
proton and neutron numbers of the $\alpha$-dripline boundaries for any
nucleon number $A$.  Similarly, it would be possible to derive
expressions for both neutron and proton driplines.

\begin{figure}
\centering
\includegraphics[width=1\columnwidth]{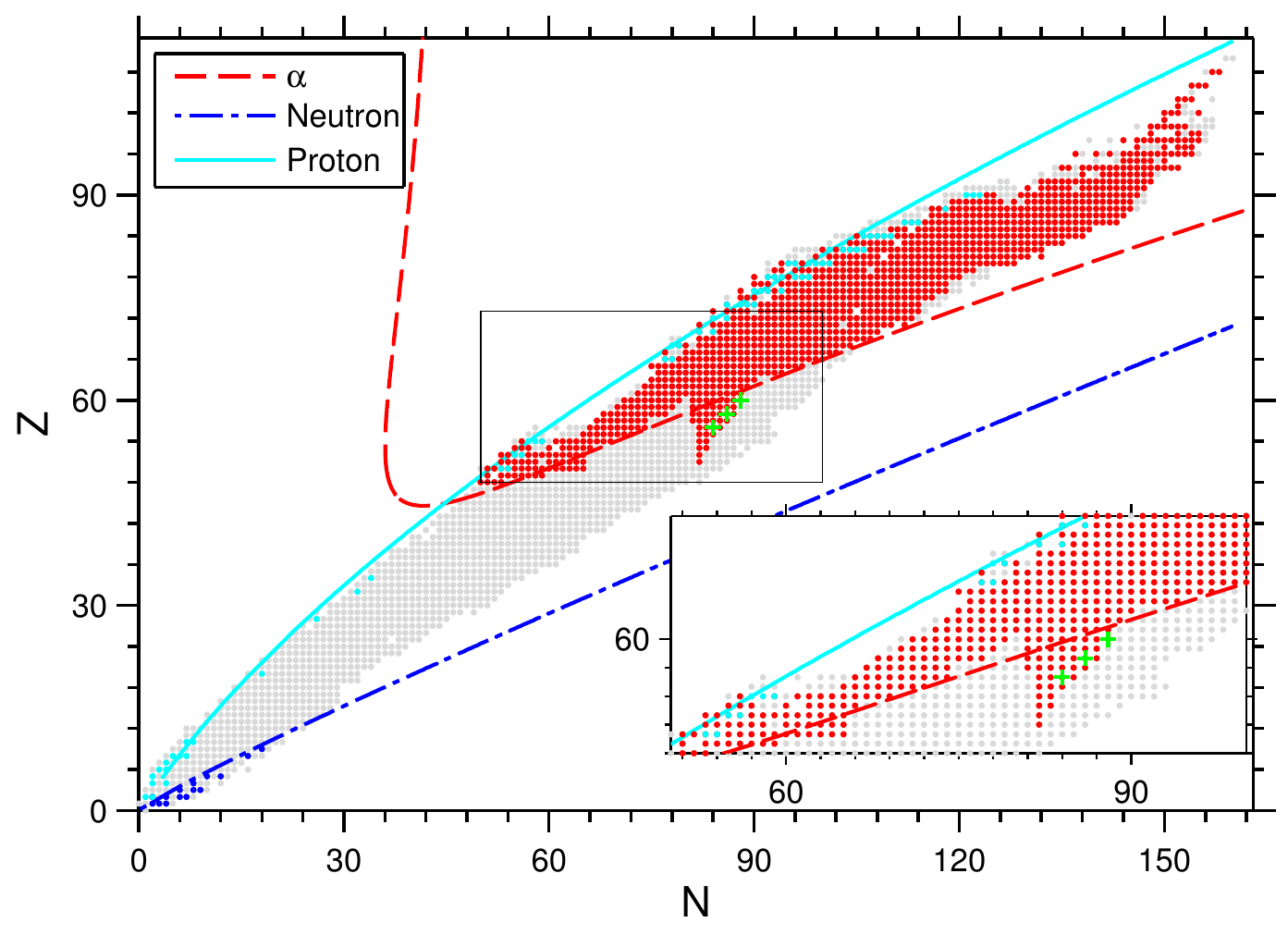}
\caption{(Colour online) The theoretical $\alpha$, neutron, and proton
  driplines in the chart of nuclei, where nuclei with a negative
  separation energy have been marked in a corresponding colour. Marked
  in green are $^{140}\text{Ba}$, $^{144}\text{Ce}$,
  and $^{148}\text{Nd}$, which represents nuclei of interest in the $\alpha$-dripline region. \label{fig nucl}}
\end{figure}

The results are shown in the nuclear chart in Fig.~\ref{fig nucl},
where nuclei with experimentally known negative neutron, proton, or
$\alpha$-separation energies are marked.  The computed curves are in
overall agreement with the measured results, which is to be expected since the
liquid drop parameters are adjusted to achieve this goal.  The
negative neutron and proton binding energies are outside the
corresponding driplines.  On the other hand, the negative
$\alpha$-bindings occur between the legs of the two solutions as
emphasized by the many known $\alpha$-unstable heavy nuclei marked in
red.

If an isotope is marginally on the unstable side of a dripline, it can
rather likely form a Borromean system by adding two identical
particles (neutrons, protons, or $\alpha$'s) to the corresponding
core-nucleus.  This is often observed for neutrons, but the repulsive
Coulomb interaction may sometimes invalidate this expectation for
protons and $\alpha$-particles.  Thus, these nuclei are promising
candidates for ground state cases of Borromean two-$\alpha$ systems.  It
is perhaps worth emphasizing that going away from the $\alpha$-driplines
into either $\alpha$-unbound or $\alpha$-bound nuclei would correspond to
$\alpha$-core ground state structures of negative or positive binding
energy, respectively.

Then the red nuclei, between the legs of the $\alpha$-dripline curves,
should be simulated by a positive $\alpha$-core energy even for the ground
state. Vice versa, outside the region this two-body energy should be
negative, and excited states of two-body energy just above zero are
the strong candidates for the $\alpha$-cluster states we are going to
investigate.  In other words, both positive and negative two-body
energies are worth investigating, both as ground states and as excited
states.

From Fig.~\ref{fig nucl} we can also deduce which mixed species of
nucleons and $\alpha$-particles are most promising in connection with
formation of Borromean states and the related $\alpha$-clusterization.
First we notice that the neutron-proton-core system is excluded as a
Borromean state due to the bound deuteron.  Borromean states involving
nucleons in general only occur for ground states close to the
corresponding nucleon driplines.  This excludes neutron-$\alpha$-core
systems since the neutron and $\alpha$ driplines never intersect or come
close to each other.  When the neutron-core is unbound the $\alpha$-core
system is bound.

In contrast, proton-$\alpha$-core systems are possible Borromean systems
along the proton dripline for systems heavier than about $N \approx
40$ or $Z > 45$.  This is especially promising in the region where
proton and $\alpha$ driplines intersect each other as shown in
Fig.~\ref{fig nucl}.  These structures are interesting and should be
investigated in details in the future. It would involve both
proton-core and $\alpha$-core effective two-body interactions.  At
present we only emphasize that this experimentally accessible region
probably would present a series of such Borromean systems.  In
addition, we notice that similar cluster structures may appear as
excited states in lighter nuclear systems.

\section{Method and parameter choices  \label{sec 2p}}

The three-body calculations require two-body potentials between the
three pairs of particles.  For $\alpha$-$\alpha$-core systems we only need
to specify the $\alpha$-$\alpha$ and $\alpha$-core potentials.  We treat all
particles as point-like and the finite sizes must then be accounted
for through effective potentials.  This also implies that the actual
choice of potentials is less important.  We can use the
energy as the crucial parameter and measure lengths relative to the
radius of the core-nucleus.  Conclusions from individual test cases
are then more general.  After definitions of the two-body potential we
define notation and principal quantities in a brief sketch of the
three-body method.

\subsection{Two-body potentials}

First we choose the $\alpha$-$\alpha$ potential as the $d$-version of the
Ali-Bodmer potential \cite{ali66} as used previously in many
applications \cite{gar13}.  This potential is angular momentum
dependent without bound states while reproducing the low-energy phase
shifts very well.  The measured energy, $0.091$~MeV, is reproduced,
and the root-mean-square radius of the corresponding resonance is
calculated to be $\langle r^{2} \rangle^{1/2} = 5.95 \,
\si{\femto\meter}$.

The second potential between $\alpha$-particle and core must necessarily
be phenomenologically adjusted. At the $\alpha$-dripline the binding
energy has to be vanishingly small, but not necessarily of the lowest
state. In the present work the antisymmetry between nucleons in the core and in the $\alpha$-particles are only accounted for by use of a shallow effective $\alpha$-core potential or by exclusion of the deepest-lying bound states. Thus, the Pauli principle is approximately obeyed by occupying only states with very small binding energy. As shown in \cite{gar97}, where the halo nuclei $^6$He and $^{11}$Li are investigated, for weakly bound systems a shallow potential and a deep potential holding Pauli forbidden states give rise to similar three-body structures provided that both potentials have the same low-energy properties. We assume the intrinsic core-spin is zero and the total angular
momentum is then exclusively from the orbital part. This implies that
the potential is central and the decisive ingredient is the radial
shape with corresponding size and strength.  The natural choice for
medium-heavy nuclei is the Woods-Saxon potential, $V(r)$, with a
constant central value and exponential fall off at larger radii, that
is
\begin{align}
V(r) = -V_0 \left( 1 + \exp\left( \frac{r - R}{a} \right) \right)^{-1} \label{eq WS} \; ,
\end{align}
where we use the diffuseness, $a = 0.65$, $R(A) = r_0 A^{1/3} +
r_{\pi} + R_{\alpha}$, $r_0 =1.16$~fm, $r_{\pi}=1.4$~fm,
$R_{\alpha}=1.7$~fm. For $A=140$ we arrive at $R = 9.1$~fm which shall
be used throughout this paper.  The remaining parameter is the
strength, $V_0$, which is tuned to give the desired energies
for any choice of angular momentum and parity.  The Coulomb potential
is for a homogeneous sharp cut-off distribution of core-charge,
$Z=56$ and $\alpha$-charge $2$, with the resulting cut off radius, $R_C=7.4$~fm.
This somewhat increased radius accounts for the finite sizes of both
core and $\alpha$-particle when a point-particle in a potential is
assumed.

\begin{figure}
\centering
\includegraphics[width=1\columnwidth]{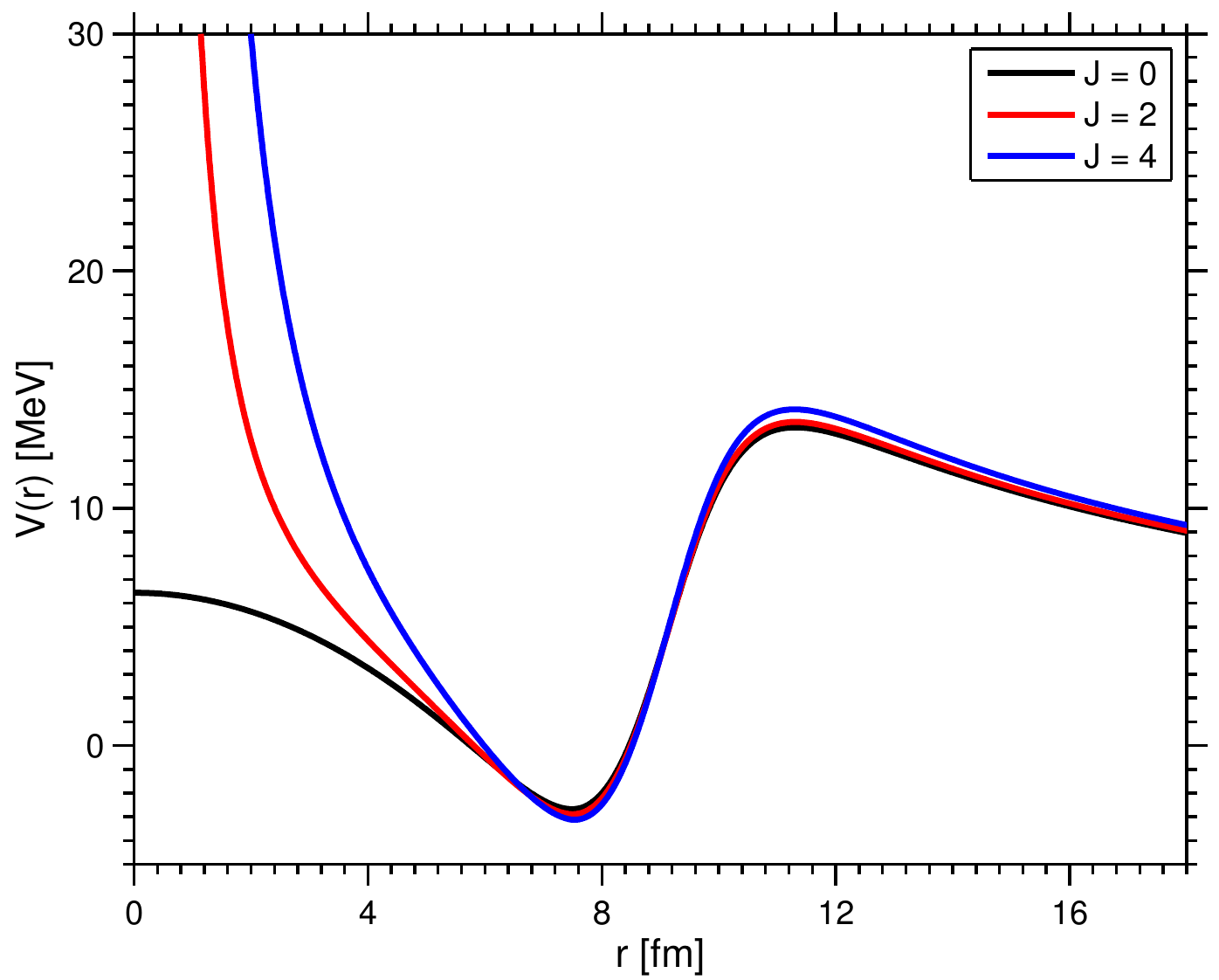}
\caption{(Colour online) The two-body potentials for three different
  angular momenta (and corresponding potential depths). The potential
  depths are adjusted to give roughly the same energy as
  reflected by very similar minima.  The depth of the interaction
  potential can be found in Table \ref{tab 2p}. \label{fig 2p pot}}
\end{figure}

The variation of angular momentum allows us to investigate the influence of
non-vanishing partial waves on the total three-body structure. It also
allows comparison between ground and excited states of both the same
and different angular momentum quantum numbers. Excited states from
the present effective potential may be unavoidable when deeper-lying
levels are occupied and Pauli-forbidden.  The effect of changing the
angular momentum can be seen in Fig.~\ref{fig 2p pot}, where the
potential depth in each case is adjusted to produce the same
energy.  The minima are then almost identical, and the only difference
is the centrifugal barrier term deviating strongly at small distance.

\begin{figure}
\centering
\includegraphics[width=1\columnwidth]{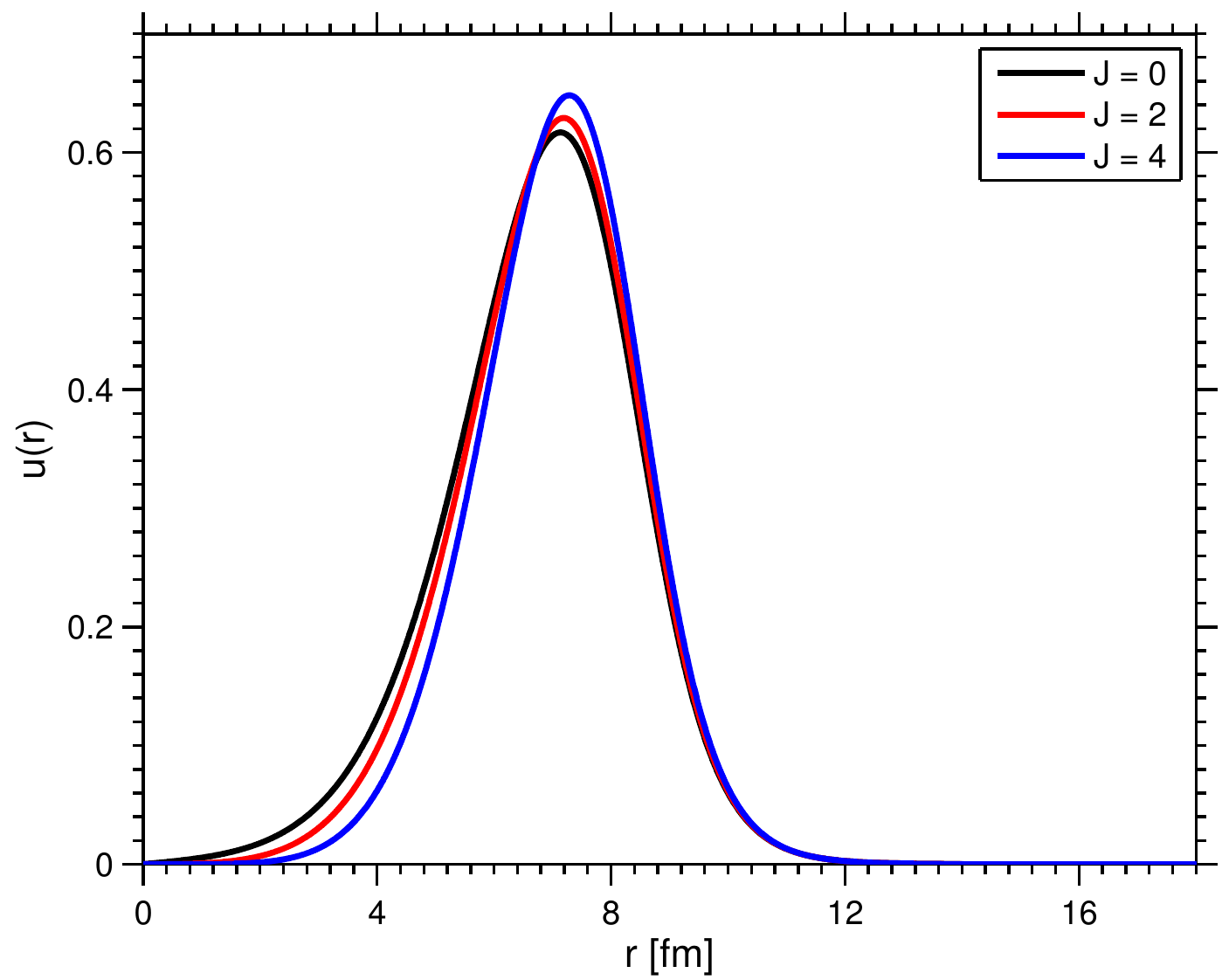}
\caption{(Colour online) The two-body radial wave functions, $u$,
  for three different angular momenta (and corresponding potential
  depths). \label{fig 2p w}}
\end{figure}

The reduced radial ground state wave functions, $u(r)$, corresponding
to the potentials in Fig.~\ref{fig 2p pot} are shown in Fig.~\ref{fig
  2p w}.  They are all spatially very similar and confined to a rather
narrow region around the common minimum.  This behaviour is unlike that
of a core and a neutron which typically has a very wide spatial extent
with a slowly decreasing tail \cite{jen04}.  The narrow distribution
is a direct consequence of the potentials in Fig.~\ref{fig 2p pot},
where the $\alpha$-particle is confined by two very steep barriers on
both sides of the minimum.  

It is especially worth noting that the Coulomb interaction for
$s$-waves is finite at $r=0$ as obtained by a homogeneous charge
distribution within a sphere.  The finite repulsion is still
sufficient to push the wave function out to the surface almost
precisely as for finite angular momenta with the additional diverging
short-range repulsion.  This angular momentum independent result is
only achieved with a relatively large charge on the core-nucleus as
for medium heavy or heavy nuclei.

\begin{table}
\centering
\caption{The two body energies and average sizes with a Ba-140
  core and an $\alpha$-particle, where the angular momentum, $J$, and
  potential depth, $V_0$, have been varied.  Positive energies, $E$, correspond to unbound states with negative binding.  We
  denote ground, first excited, second excited, and third excited
  states by (G), (F), (S) and (T), respectively.  Energies are in
  $\si{\mega\electronvolt}$, and distances are in
  $\si{\femto\meter}$. \label{tab 2p}}
\begin{ruledtabular}
\begin{tabular}{ *{5}{c}} 
$V_0$    &   $J$    &  State   &  $E$       &  $\langle r^2 \rangle^{1/2}$ \\
\colrule
26.25    &   $0$    &   G.     &  0.1      &     7.0       \\
         &          &   F.     &  4.3      &     5.9        \\
         &          &   S.     &  7.4      &     5.4        \\
\hline
27.1     &   $2$    &   G.     &  0.0      &     7.1    \\
         &          &   F.     &  4.9      &     6.4     \\
         &          &   S.     &  9.1      &     6.4  \\
\hline
28.8     &   $4$    &   G.     &  0.1      &     7.2   \\
         &          &   F.     &  5.7      &     6.8   \\
\end{tabular}
\end{ruledtabular}
\end{table}

The adjusted depths corresponding to $J = 0, 2,$ and $4$ are collected
in Table \ref{tab 2p}. The resulting $\alpha$-core energies and
the root-mean square radii are also given in this table for ground and
all excited states. Due to the large barriers around the minimum
these unbound states of positive energy are sufficiently well defined
to allow computation of their radii. The potential depths are adjusted with angular momentum specifically to compensate the centrifugal barrier and leave the energy of the weakest bound state at essentially the same value. The intent is to have a slightly unbound two-body subsystem in the three-body system. The deepest potential will be used as the $\alpha$-core in the three-body system, as both the effect of changing angular momentum and the low lying excited states are of interest. Using the potential depth associated with $J=4$ in a calculation with $J=0$ may clearly produce
more bound states. The radii of the states in Table \ref{tab 2p}
increase with increasing energy, but only moderately. The
actual values would be strongly dependent on the radius of the
potential. The peaks in Fig.~\ref{fig 2p w} occur at around $7.2$~fm
which is about the size of the core plus $\alpha$-particle charge
radius. Thus fortunately, but not surprisingly, the $\alpha$-particle is
located at the surface of the core.

\subsection{Three-body formalism \label{sec 3p}}

The three-body calculations are carried out by use of the adiabatic
hyper-spherical expansion method \cite{nie01}.  First the Jacobi
coordinates are defined as mass scaled vectors, ${\bf x}$ and ${\bf
  y}$, between one pair of particles, and between their center-of-mass and
the third particle respectively.  The relative orbital angular momenta, $l_x$ and
$l_y$, are related to this choice of Jacobi coordinates.  Three
different choices are possible.  The hyperspherical coordinates are
defined by the hyperradius, $\rho$, and five hyperangles, $\Omega$.
The (coordinate independent) definition of $\rho$ involves an
arbitrary normalization mass, $m$, which has no influence on the
result and is only used for notational convenience.  

We first solve the hyper-angular part of the Faddeev equations for
fixed average radius $\rho$.  Each partial wave in each Faddeev
component is expanded on the set of Jacobi polynomials from constants
to the highest order defined by $K_{max}$.  This provides a set of
angular eigenvalues, $\lambda_n(\rho)$, and eigenfunctions,
$\Phi_n(\rho,\Omega)$, where all quantities depend on $\rho$.  The
solution to these equations produce the effective potentials
\begin{align}
V_{eff}(\rho) = \frac{\hbar^{2}}{2 m} \left( \frac{\lambda_n(\rho) + 15/4}{\rho^{2}} \right), \label{eq eff pot}
\end{align}
where $\lambda_n(\rho)$ is the crucial ingredient.  The total wave
function, $\Psi$, is expanded on the complete set, $\Phi_n$,
\begin{align}
\Psi = \displaystyle\sum_n \rho^{-5/2} f_n(\rho) \Phi_n(\rho,\Omega),
\label{wavef}
\end{align}
where $f_n$ are the hyper-radial wave functions.  They are determined
by the coupled set of hyperradial equations arising from insertion of
$\Psi$ into the Faddeev equations, that is
\begin{align}
&\left( - \frac{\partial^{2}}{\partial \rho^{2}} + \frac{1}{\rho^{2}} \left( \lambda_n(\rho) + \frac{15}{4} \right) - Q_{nn} - \frac{2m E}{\hbar^{2}} \right) f_n(\rho),  \notag \\ & =  \label{eq rad}
\displaystyle\sum_{n^{\prime}\neq n} 
\left( 2 P_{n n^{\prime}} \frac{\partial}{\partial \rho} + Q_{n n^{\prime}} \right) f_{n^{\prime}}(\rho) \; ,
\end{align}
where $E$ is the energy. The coupling terms, $P$ and $Q$, are given by
\begin{align}
P_{nn^{\prime}}(\rho) 
&= \left\langle \Phi_n \left\vert \frac{\partial}{\partial \rho} \right\vert \Phi_{n^{\prime}} \right\rangle_{\Omega}, \\
Q_{n n^{\prime}} (\rho) 
&= \left\langle \Phi_n \left\vert \frac{\partial^{2}}{\partial \rho^{2}} \right\vert \Phi_{n^{\prime}} \right\rangle_{\Omega}  \;,
\end{align}
where the expectation values are over the hyperangles, $\Omega$, for fixed
$\rho$.  The convergence with the number of included adiabatic
potentials is usually very fast, and only 4-6 are necessary in
Eq.~(\ref{eq rad}).

\begin{figure}
\centering
\includegraphics[width=0.9\columnwidth]{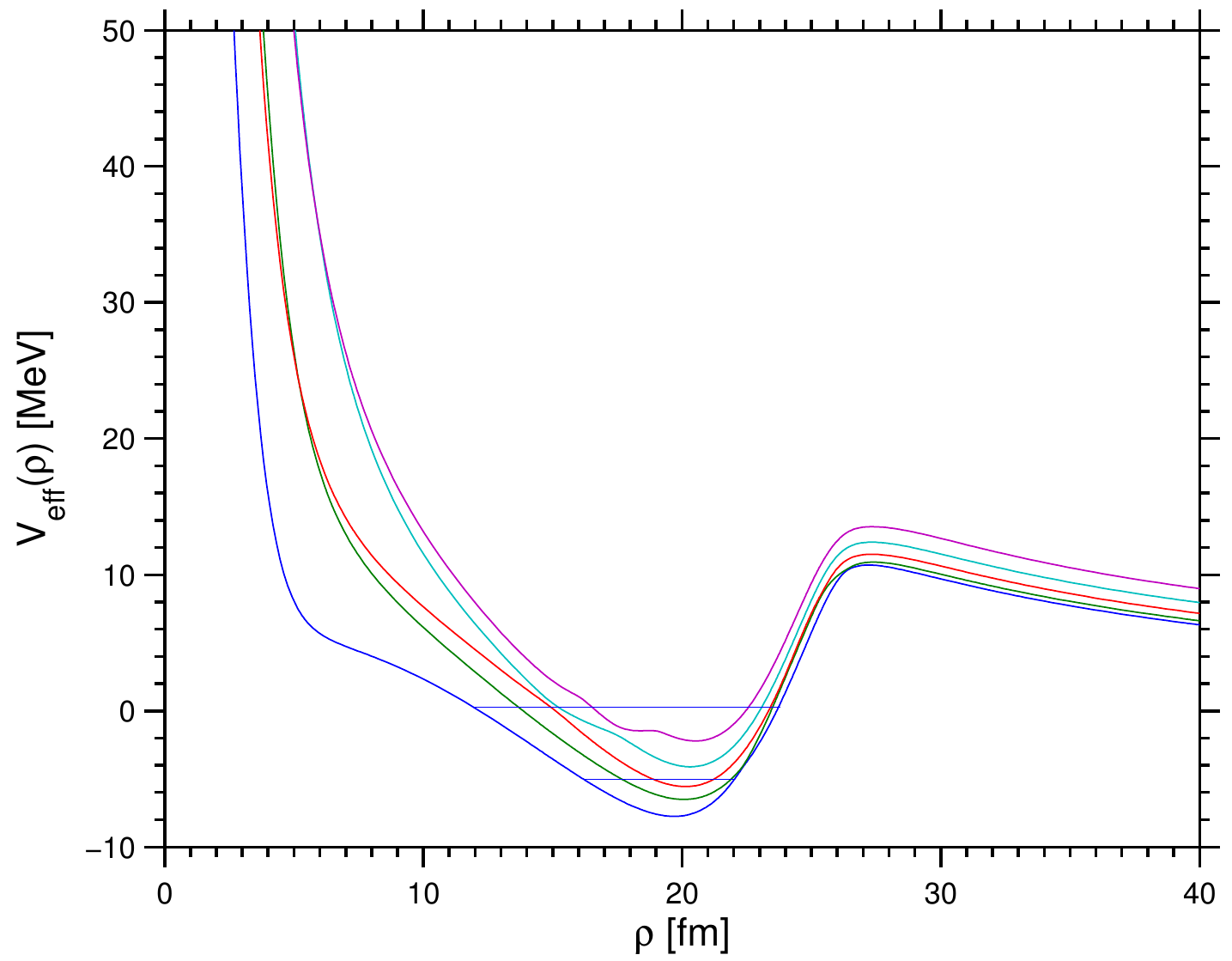}
\caption{(Colour online) Effective potentials, Eq.~(\ref{eq eff pot}),
  calculated from the five lowest $\lambda_n(\rho)$, for the case
  where $V_0 = 28.8 \, \si{\mega\electronvolt}$ and $J^{+} =
  0^{+}$. The horizontal lines correspond to the lowest, and first
  excited state of the lowest $\lambda_n(\rho)$, when it is
  approximated by a harmonic oscillator. It is noteworthy how similar
  the effective potentials are for the different $\lambda_n$
  functions. \label{fig pot}}
\end{figure}

\section{Core plus two-alpha properties \label{sec nd148}}

The previous section introduced the two-body potentials and the three-body formalism, which will be applied in the present section. Here $^{148}\text{Nd}$ is considered as a three body system consisting of a $^{140}\text{Ba}$ core and two $\alpha$ particles. These nuclei are chosen as they are at the edge of the $\alpha$ unstable region in Fig.~\ref{fig nucl}. The purpose of this section is to study the nature of a general, relatively heavy, $\alpha$-Borromean system. As the same potential will be used for all partial waves, and as this potential was only adjusted to create a slightly unbound two-body system, energy levels cannot be expected to be reproduced. Of particular interest are the distributions among both the effective potentials and the partial waves, as well as the spatial distributions. In Sec.~\ref{sec ba142} a detailed fine tuning of the individual partial waves is included for a similar system ($^{142}\text{Ba}$ consider as $^{134}\text{Te}+\alpha+\alpha$) to reproduce both energy levels and electric transition probabilities.

The calculations presented here treat the core as an inert particle with angular momentum and parity $0^+$. Therefore, the effects arising from excitations of the $^{140}$Ba core, in the $^{148}$Nd case, or the $^{134}$Te core, in the $^{142}$Ba case, into the 2$^+$ excited state (at 0.60 MeV in $^{140}$Ba and 1.28 MeV in $^{134}$Te) will not be considered.

The solutions are obtained in two steps.  First the angular
wave functions are calculated and second we solve the coupled radial
set of equations.  In the first step both angular wave functions, and
radial potentials, and their couplings are produced.  In the first
subsection, we discuss the properties of these solutions, and in the
second subsection we present the radial structure in simple geometric
terms.

\subsection{Angular three-body structure}

The five lowest of the effective potentials given in Eq.~(\ref{eq eff pot})
are shown in Fig.~\ref{fig pot} for angular momentum and parity,
$0^+$, and a given appropriate strength, $V_0=28.8$~MeV, of the
Woods-Saxon potential.  The lowest minimum value is about $-8$~MeV and
located close to $\rho \approx 20$~fm.  The potential has a rather
steep barrier rising to about $+11$~MeV at roughly $27$~fm, after which
it decreases slowly towards zero as $\rho$ increases to infinity. The
fall-off is proportional to $1/\rho$ because the Coulomb potentials
are responsible for this long-range behaviour.  This implies
proportional fall-off for all potentials, since the large-distance
Coulomb interactions are the same for all adiabatic potentials.  The
increase of the potential for small $\rho$ is due to the centrifugal
barrier behaviour of $1/\rho^2$, while the Coulomb potentials remain
finite through the assumption of homogeneous charge distributions.

The higher-lying adiabatic potentials are remarkably similar to the
lowest and each only shifted by about $1.5$~MeV, very crudely
independent of $\rho$.  The zero point motion of the best fit of the
lowest potential by a one-dimensional oscillator is about $2.7$~MeV
($\hbar\omega \approx 5.5$~MeV).  The first excited oscillator energy
is then at about $8.5$~MeV above the oscillator bottom.  The shifted
zero point in our potential is at about $-8$~MeV producing two
oscillator estimates at about $-5.0$~MeV and $+0.5$~MeV as indicated
by the horizontal lines in Fig.~\ref{fig pot}.

The distance between neighbouring adiabatic potentials is roughly about
$1.5$~MeV.  The first excited state in the lowest potential at about
$+0.5$~MeV is then at about the same position as the ground state of
the fourth potential, which is estimated to be at about $4\times 1.5$~MeV
$+2.7$~MeV above the lowest minimum at $-8$~MeV, that is $+0.7$~MeV.
This implies that the wave functions of the lowest-lying two states in
the $0^+$ spectrum can be expected almost entirely built on individual
potentials, unless of course the couplings between the adiabatic
potentials are unusually strong.  The third excited state could
energy-wise instead be composed of comparable components from first
and fourth potentials.

Higher angular momentum potentials are rather similar but with minima
shifted upwards by the centrifugal barrier amounting to roughly
$0.4$~MeV and $1.4$~MeV for $2^+$ and $4^+$, respectively.  The large
distance behaviour is essentially maintained, whereas the increase and
eventual divergence at short distance accelerate with angular momentum
as in Fig.~\ref{fig 2p pot}.  Finally, modest variation of the
two-body potential strength will only displace the curves slightly,
and the effect is most noticeable at large distances, where it has no
effect on the bound state structure.

\begin{table}
\centering
\caption{The three body energies of $^{148}\text{Nd}$ $(^{140}\text{Ba} + \alpha + \alpha)$, as well as average
  $\alpha$-$\alpha$ and $\alpha$-core distances for different angular momenta
  with $V_0=28.8$~MeV, $a=0.65$~fm, $R=9.1$~fm, $R_C=7.4 $~fm.  The
  weights of the contributing adiabatic potentials are given for each
  state in the last five columns.  All energies are in
  $\si{\mega\electronvolt}$ and all distances are in
  $\si{\femto\meter}$. \label{tab overview}}
\begin{ruledtabular}
\begin{tabular}{ l *{3}{c} *{5}{c}} 
  & & & &   \multicolumn{5}{c}{Weights of potentials $(\%)$} \\
 \cline{5-9}
  $J^{\pi}$    & $E$  &  $\langle r_{\alpha \alpha}^2 \rangle^{\frac{1}{2}}$  &   $\langle r_{\alpha c}^2 \rangle^{\frac{1}{2}}$  & 1 & 2 & 3 & 4 & 5  \\
\colrule
 $0^{+}$   &  -4.9    &    4.8     &    6.9  &   95   &  4   &  0   &   1   &   0   \\
  $0^{+}$  &  -3.7    &    12.0    &    7.0  &   7    &  92  &  1   &   0   &   0   \\
  $0^{+}$  &  -2.6    &    10.9    &    7.0  &   3    &  1   &  94  &   2   &   0   \\
   $0^{+}$ &  -0.8    &    10.2    &    7.2  &   17   &  3   &  2   &   74  &   4   \\
\hline
   $2^{+}$ &  -4.5    &    4.7     &    7.0  &   95   &  5   &  0   &   0   &   0   \\
   $2^{+}$ &  -3.3    &    12.0    &    7.0  &   7    &  89  &  3   &   1   &   0   \\
   $2^{+}$ &  -2.3    &    9.6     &    7.0  &   2    &  4   &  22  &  70   &   2   \\
   $2^{+}$ &  -0.1    &   10.9     &    7.4  &   2    &  1   &   9  &  22   &  66   \\
\hline
   $4^{+}$ &  -3.5    &    4.9     &    7.0  &   95   &  4   &  0   &   0   &   0   \\
   $4^{+}$ &  -2.4    &    12.4    &    7.1  &   7    &  84  &  9   &   0   &   0   \\
   $4^{+}$ &  -1.6    &    9.7     &    7.0  &   1    &  10  &  81  &   6   &   2   \\
   $4^{+}$ &  -0.7    &    8.8     &    7.0  &   0    &  1   &  10  &   78  &   10  \\
\end{tabular}
\end{ruledtabular}
\end{table}

The calculated energies are given in Table \ref{tab overview} for the
potential strength $V_0=28.8$~MeV and different angular momenta.
Decreasing the attraction to $V_0=26.25$~MeV only one $0^+$ state
($2^+$, $4^+$) is bound at $-0.2$~MeV. A further increase of strength
to $V_0=27.1$~MeV provides two bound $0^+$ states at $-1.8$~MeV and 
$-0.6$~MeV, and two bound $2^+$ states at $-1.3$~MeV and $-0.2$~MeV.  When
$V_0=28.8$~MeV, corresponding to Fig.~\ref{fig pot}, we find four
bound state solutions for each set of quantum numbers, $0^+$, $2^+$
and $4^+$.  Thus, the three-body bound states appear much faster and
more abundantly than the two-body $\alpha$-core potentials in
Fig.~\ref{fig 2p pot}.  The oscillator estimate of about $-5.0$~MeV
and $0.5$~MeV for the two lowest $0^+$ states built on the lowest
adiabatic potential is rather accurate as only one corresponding bound
state appears at $-4.9$~MeV, see Table \ref{tab overview}. Unbound resonance states are not computed until Sec.~\ref{sec ba142}, where $^{142}\text{Ba}$ is examined in detail.

The structure of these states is known through the calculated
properties of the wave functions. We consider first the contributions
from the different adiabatic potentials to the individual states.
Only five potentials are necessary to ensure accurate radial
solutions.  The relative weights in Table \ref{tab overview} are
remarkably simple with one entirely dominating potential for all wave
functions.  The two weakest bound $2^+$ states are the most
fractionated with a division of ($22\%$, $70\%$) and ($22\%$,$66\%$)
on third and fourth, and fourth and fifth potentials, respectively.

In general, each potential then essentially carries the full weight of
a given state, such that the lowest potential corresponds to the
lowest energy, the second potential and the second lowest energy are
related, etc.  This confirms the main conclusion of one adiabatic
potential per state obtained from the estimate by use of an oscillator
approximation without couplings between potentials.  The third excited
$0^+$ state begins to have contributions from both first and fourth
adiabatic potentials.  The excitation on the lowest potential competes
with the lowest energy on the fourth potential, and two configurations
arise.  The second excited $2^+$ state is fractionated between third
and fourth potentials, now because the potentials happen to be
rather close-lying and the couplings are therefore more effective.

\begin{table}
\centering
\caption{The weights of each partial wave for the potential specified
  in Table \ref{tab overview}. Here $l_x$ denotes the relative angular
  momentum between the two particles, and $l_y$ denotes
  the angular momentum of the third particle relative to the center of
  mass of the first two particles. The fifth column give the order,
  $K_{max}$, of the Jacobi polynomium used for the corresponding
  partial wave.  The last four columns each describe ground (G), first
  (F), second (S) and third (T) excited states.  Components where all
  states have a weight less than $0.04$ are omitted.  \label{tab 2p 4} }
\begin{ruledtabular}
\begin{tabular}{c| *8{c}}
$J^{\pi}$   &     Jacobi         & $l_x$ &   $l_y$      & $K_{max}$ &     G.    &   F.       &   S.   &    T.   \\
\colrule
$0^{+}$    & $\alpha$-$\alpha$   &   0   &   0          &    80     &  0.79     &    0.73    &  0.72   &  0.72  \\
           &                     &   2   &   2          &    60     &  0.19     &    0.22    &  0.22   &  0.22  \\
           &                     &   4   &   4          &    50     &  0.02     &    0.04    &  0.05   &  0.05  \\
\hline
$0^{+}$    & $\alpha$-c          &   0   &   0          &    100    &  0.43     &    0.51    &  0.04   &  0.07  \\
           &                     &   1   &   1          &    80     &  0.39     &    0.48    &  0.09   &  0.06  \\
           &                     &   2   &   2          &    60     &  0.10     &    0.01    &  0.85   &  0.01  \\
           &                     &   3   &   3          &    50     &  0.00     &    0.00    &  0.00   &  0.79  \\
\hline
\hline
$2^{+}$    & $\alpha$-$\alpha$   &   0   &   2          &    70     &  0.82     &    0.07    &  0.26   &  0.04  \\
           &                     &   2   &   0          &    70     &  0.05     &    0.72    &  0.45   &  0.54  \\
           &                     &   2   &   2          &    50     &  0.05     &    0.08    &  0.16   &  0.29  \\
           &                     &   2   &   4          &    40     &  0.07     &    0.02    &  0.06   &  0.01  \\
           &                     &   4   &   2          &    40     &  0.01     &    0.09    &  0.05   &  0.05  \\
           &                     &   4   &   4          &    30     &  0.00     &    0.01    &  0.01   &  0.05  \\
\hline
$2^{+}$    & $\alpha$-c          &   0   &   2          &    70     &  0.18     &    0.25    &  0.02   & 0.00   \\
           &                     &   2   &   0          &    70     &  0.19     &    0.23    &  0.03   & 0.00   \\
           &                     &   1   &   1          &    50     &  0.39     &    0.50    &  0.03   & 0.00   \\
           &                     &   2   &   2          &    50     &  0.06     &    0.01    &  0.05   & 0.23   \\
           &                     &   1   &   3          &    40     &  0.04     &    0.00    &  0.38   & 0.07   \\
           &                     &   3   &   1          &    40     &  0.05     &    0.00    &  0.37   & 0.07   \\
           &                     &   3   &   3          &    40     &  0.01     &    0.00    &  0.03   & 0.35   \\
           &                     &   2   &   4          &    40     &  0.00     &    0.00    &  0.01   &  0.05  \\
           &                     &   4   &   2          &    40     &  0.00     &    0.00    &  0.02   &  0.04  \\
           &                     &   4   &   4          &    30     &  0.00     &    0.00    &  0.03   &  0.05  \\
\hline
\hline
$4^{+}$    & $\alpha$-$\alpha$   &   2   &   2          &    80     &  0.09     &    0.15    &  0.34   &  0.27  \\
           &                     &   0   &   4          &    50     &  0.78     &    0.02    &  0.07   &  0.08  \\
           &                     &   4   &   0          &    50     &  0.00     &    0.68    &  0.15   &  0.26  \\
           &                     &   4   &   2          &    48     &  0.01     &    0.03    &  0.18   &  0.02  \\
           &                     &   2   &   4          &    48     &  0.06     &    0.03    &  0.18   &  0.27  \\
           &                     &   6   &   2          &    20     &  0.00     &    0.06    &  0.02   &  0.02  \\
\hline
$4^{+}$    & $\alpha$-c          &   2   &   2          &    80     &  0.40     &    0.51    &  0.03   &  0.02  \\
           &                     &   1   &   3          &    80     &  0.20     &    0.26    &  0.02   &  0.00  \\
           &                     &   3   &   1          &    66     &  0.22     &    0.22    &  0.03   &  0.00  \\
           &                     &   2   &   4          &    68     &  0.01     &    0.00    &  0.00   &  0.20  \\
           &                     &   4   &   2          &    68     &  0.01     &    0.00    &  0.00   &  0.21  \\
           &                     &   4   &   0          &    50     &  0.04     &    0.00    &  0.41   &  0.03  \\
           &                     &   0   &   4          &    50     &  0.03     &    0.00    &  0.42   &  0.03  \\
           &                     &   4   &   4          &    50     &  0.00     &    0.00    &  0.01   &  0.06  \\
           &                     &   5   &   1          &    40     &  0.00     &    0.00    &  0.01   &  0.09  \\
           &                     &   1   &   5          &    40     &  0.00     &    0.00    &  0.01   &  0.10  \\
\end{tabular}
\end{ruledtabular}
\end{table}

The structure revealed by the partial wave decomposition of the wave
functions is seen in Table \ref{tab 2p 4}, where only the
contributions amounting to more than $4\%$ are included.  We give
decompositions in both the two different Jacobi coordinate sets. Here
it is worth noticing that only even angular momenta are allowed
between the two $\alpha$-particles due to the identical boson
characteristics.  We first emphasize that the small attractions, where
the energies approach all the way down to zero, all down to
the percent level produce the same partial wave decomposition
independent of the specific energy.  We therefore only show
the results for one strength.  The structures remain unchanged because
essentially only the potential energies are moved corresponding to a
shifted energy scale.

\begin{figure}[ht]
\centering
{\includegraphics[width=0.95\columnwidth]{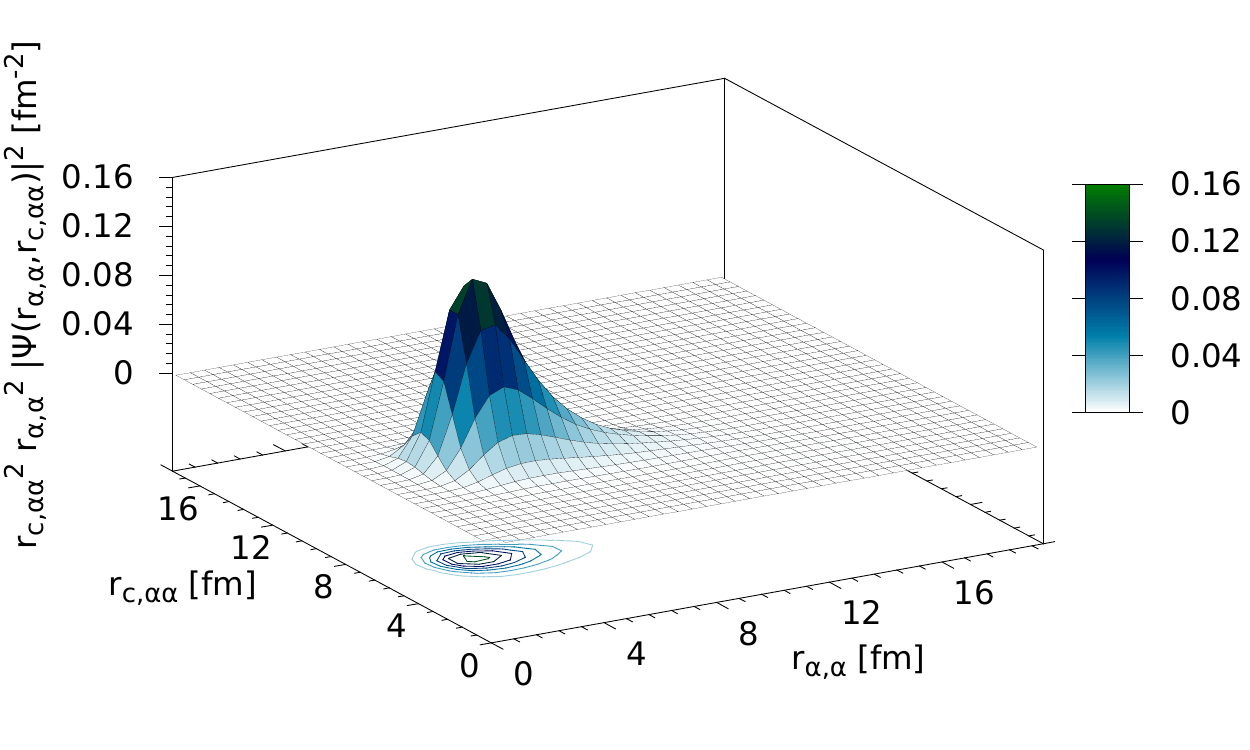}}

{\includegraphics[width=0.95\columnwidth]{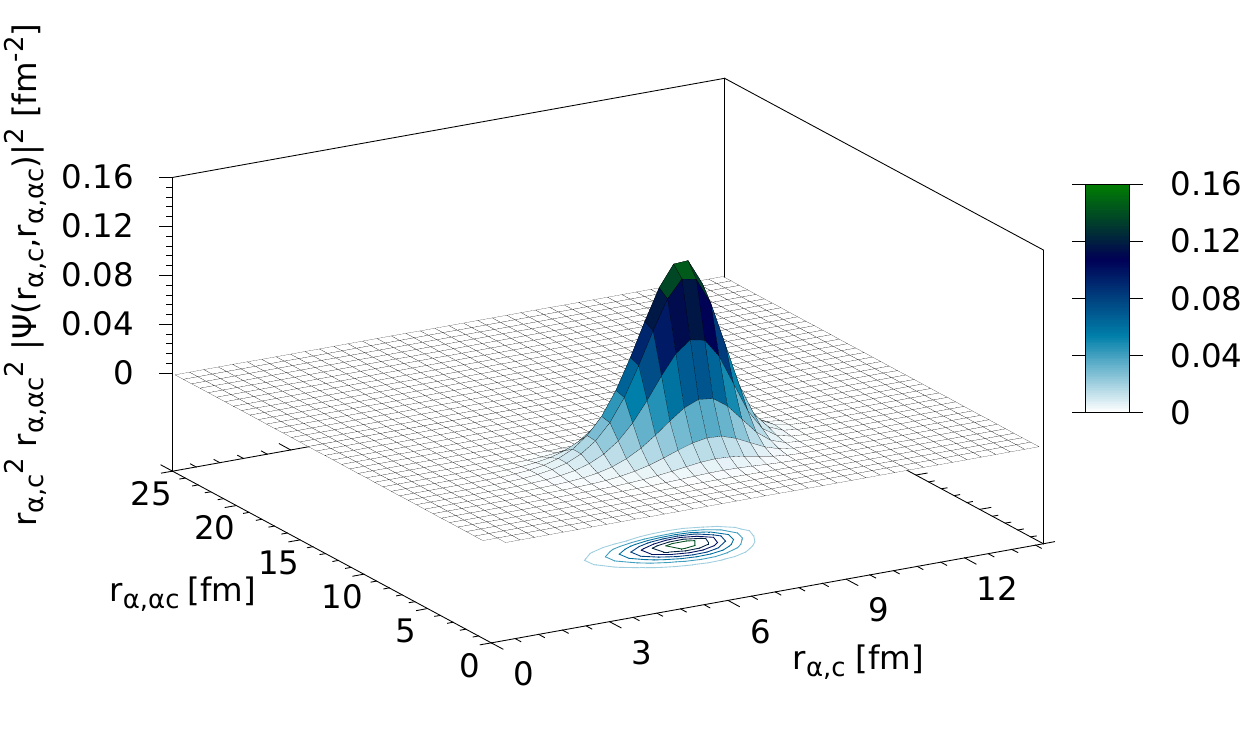}}

\caption{(Colour online) The probability distribution for the $0^+$
  ground state wave with the $\alpha$-core potential specified in Table
  \ref{tab overview}.  Projected contour curves are shown at the
  bottom of each figure.  The distance variables correspond to the two
  different Jacobi sets, where top and bottom panels are for the first
  and second Jacobi sets, respectively.  \label{fig 3d 00}}
\end{figure}

For the $0^+$ states we find more than $70\%$ of $\alpha$-$\alpha$ $s$-waves
in all solutions, while the $d$-waves absorb most of the remaining
probabilities.  This is obviously consistent with the stronger
$\alpha$-$\alpha$ attraction in the lowest partial waves.  However, the
$\alpha$-core potential may prefer another competing structure.  The two
lowest $0^+$ states have roughly equal amounts of $\alpha$-core relative
$s$- and $p$-waves, whereas the third and fourth $0^+$ states are
dominated by $d$- and $f$-waves, respectively.  In combination with
the results from Table \ref{tab overview} this indicates that these
higher-lying adiabatic potentials are dominated by $d$- and $f$-waves.

The $2^+$ states must have non-zero angular momentum partial
waves. The most favourable structure seen in Table \ref{tab 2p 4}, is
apparently $s$-waves between the two $\alpha$-particles but a
distribution for the $\alpha$-core structure of equal $s$- and $d$-waves
and twice as much $p$-waves.  In contrast, the first excited state has
dominating $\alpha$-$\alpha$ $d$-waves and comparable to the ground state
contributions from $g$-waves.  The second and third
excited states have also $\alpha$-$\alpha$ $d$-waves as the largest
components.  The $\alpha$-core contributions are now moved to roughly
equal $p$- and $f$-waves, and comparable $d$- and $f$-waves,
respectively for second and third excited state.

The $4^+$ states must have even larger finite angular momentum
contributions than the $2^+$ states.  In Table \ref{tab 2p 4} we find
again that $s$-waves dominate for ground state, while the first excited state is dominated by $g$-waves in
the $\alpha$-$\alpha$ subsystem.  In the $\alpha$-core subsystem, these two
states have the largest contributions from $d$-waves and roughly half
as much for $p$- and $f$-waves. The third and fourth $4^+$ states have
in the $\alpha$-$\alpha$ subsystem more than $50\%$ of $d$-waves and roughly
half as much $g$-waves. In the $\alpha$-core subsystem $s$- and $g$-, and
$d$- and $g$-wave components are about equal, respectively for the third
and fouth $4^+$ states. The higher-lying states receive significant
contributions from more partial waves than ground and excited states.

These rather complicated variations in structure for the different
states are dictated by minimizing the total energies. This involves
combinations of the two-body interactions which in the present cases
always prefer the lowest partial waves.  The final results are then
obtained by combining the minimization with total angular momentum
conservation, partial wave couplings, and orthogonality of all pairs of
states.

\subsection{Radial structure}

The angular eigenvalues, $\lambda_n$, provide information about the
crucial radial potentials, which in turn through Eq.~(\ref{eq rad})
determine the radial wave functions, $f_n(\rho)$.  The linear
combination with the angular parts, $\Phi_n$, from Eq.~(\ref{wavef})
gives access to complete information about each of the solutions.  We
shall in particular be concerned with probability distributions for the
$\alpha$-$\alpha$ and $\alpha$-core distances.

In Fig.~\ref{fig 3d 00} we show the probability distribution for the
$0^+$ ground state in two coordinate systems corresponding to the two
different Jacobi coordinates.  The structure is relatively simple with
only one peak at an $\alpha$-$\alpha$ distance of about $4$~fm from the top
panel and an $\alpha$-core distance of about $7$~fm from the bottom panel
of Fig.~\ref{fig 3d 00}.

\begin{figure}
\centering
{\includegraphics[width=0.95\columnwidth]{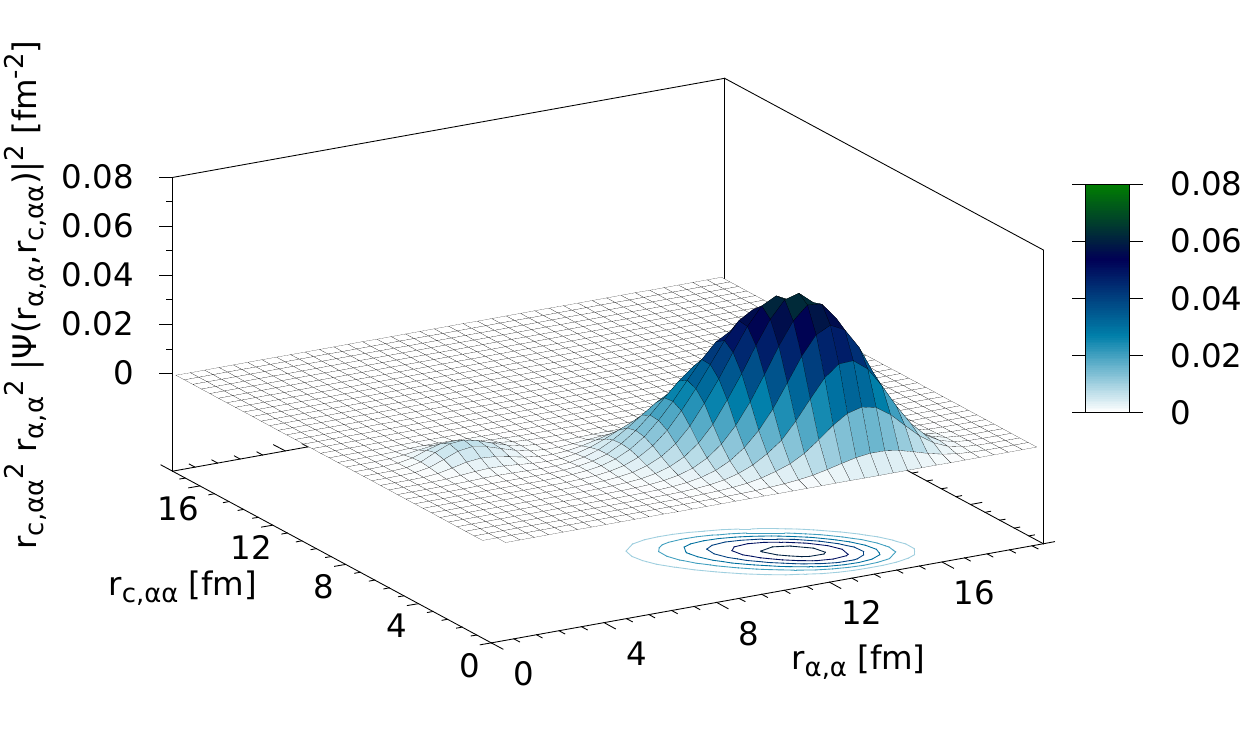}}

{\includegraphics[width=0.95\columnwidth]{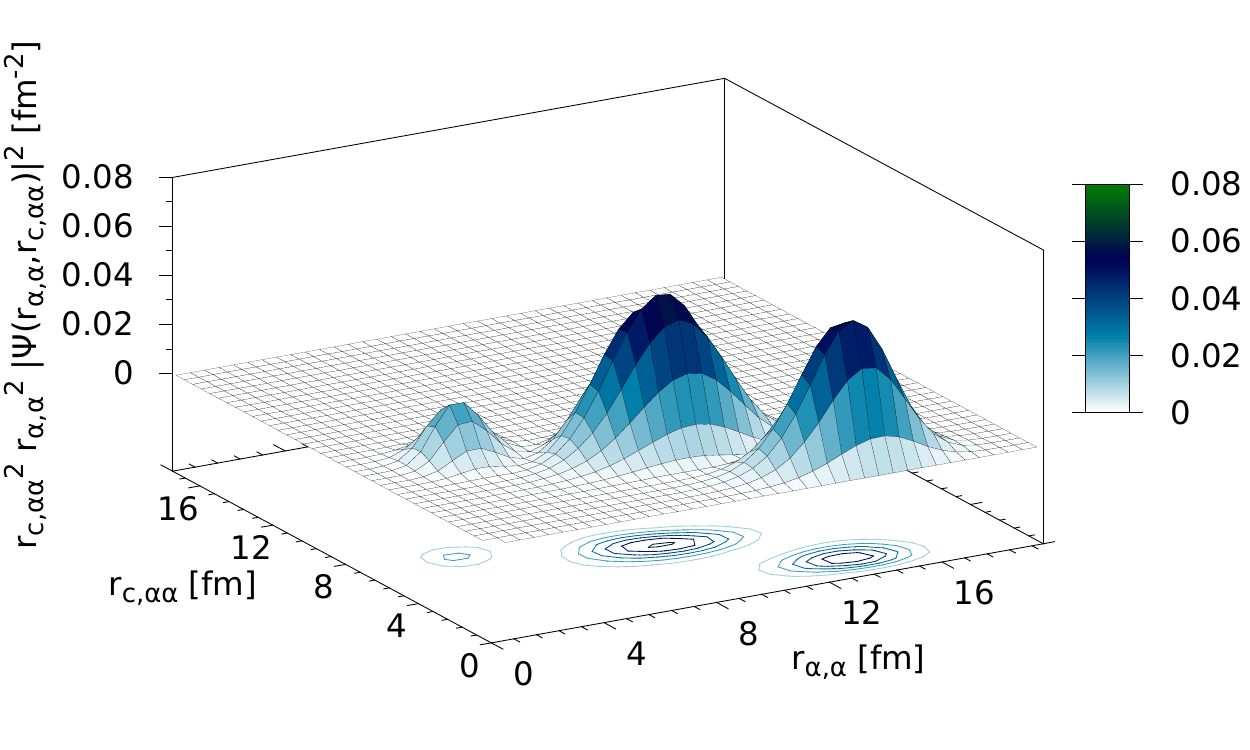}}

{\includegraphics[width=0.95\columnwidth]{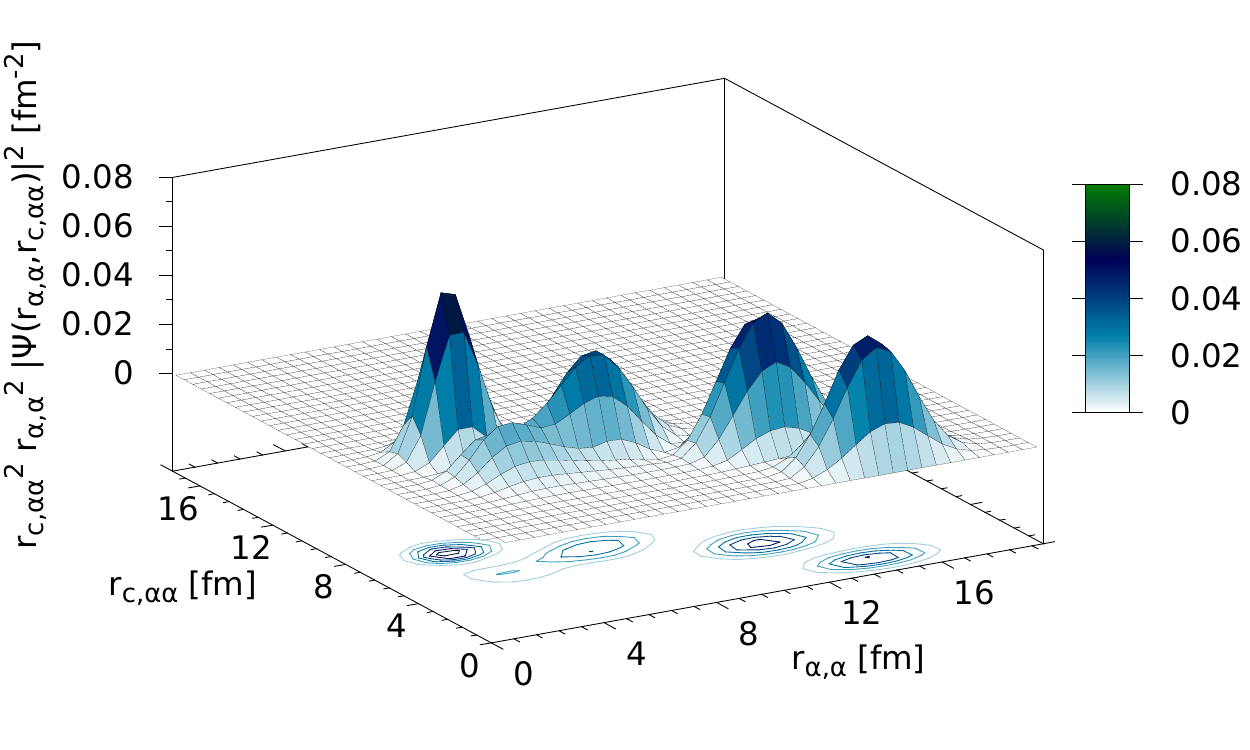}}

\caption{(Colour online) The same as top panel of Fig.~\ref{fig 3d 00}.
  for the three excited $0^+$ states.  \label{fig 3d 30}}
\end{figure}

The probability distributions for the excited $0^+$ states are shown
in Fig.~\ref{fig 3d 30} for the first Jacobi set where the
$x$-coordinate is between the two $\alpha$-particles.  In all these
excited states we find probability distributions between core and
$\alpha$-particle almost identical to that of the ground state as shown
in the lower part of Fig.~\ref{fig 3d 00}.  Consequently, we do not
show these distributions.  However, the identical distributions
demonstrate that the $\alpha$-particles strongly prefer to be located at
the surface of the core as for the isolated two-body system with the
wave function shown in Fig.~\ref{fig 2p w}.  The reason is that the
$\alpha$-core potential overrules all other possible effects when
determining the three-body structure.  Apparently the $\alpha$-$\alpha$
potential is strongly attractive and the potential energy minimum is
rather narrow and deep.
 
However, the $\alpha$-$\alpha$ distribution varies from top to bottom in
Fig.~\ref{fig 3d 30}, and they are also different from the ground
state distribution.  The first excited state shows a broader
distribution around the $\alpha$-$\alpha$ distance of $13$~fm with a marginal
reminiscence of a peak at the ground state location of about $4$~fm.
The second excited state has three peaks at $\alpha$-$\alpha$ distances of
about $14$~fm, $9$~fm and $4$~fm.  The third excited $0^+$ state
continues the trend by containing four peaks at $\alpha$-$\alpha$ distances of
$14$~fm, $12$~fm, $7$~fm, and $4$~fm.  These different structures
reflect the different structures of the corresponding adiabatic
potentials, which deliver the dominating contributions to each of the
excited states.

\begin{figure}
\centering
{\includegraphics[width=0.95\columnwidth]{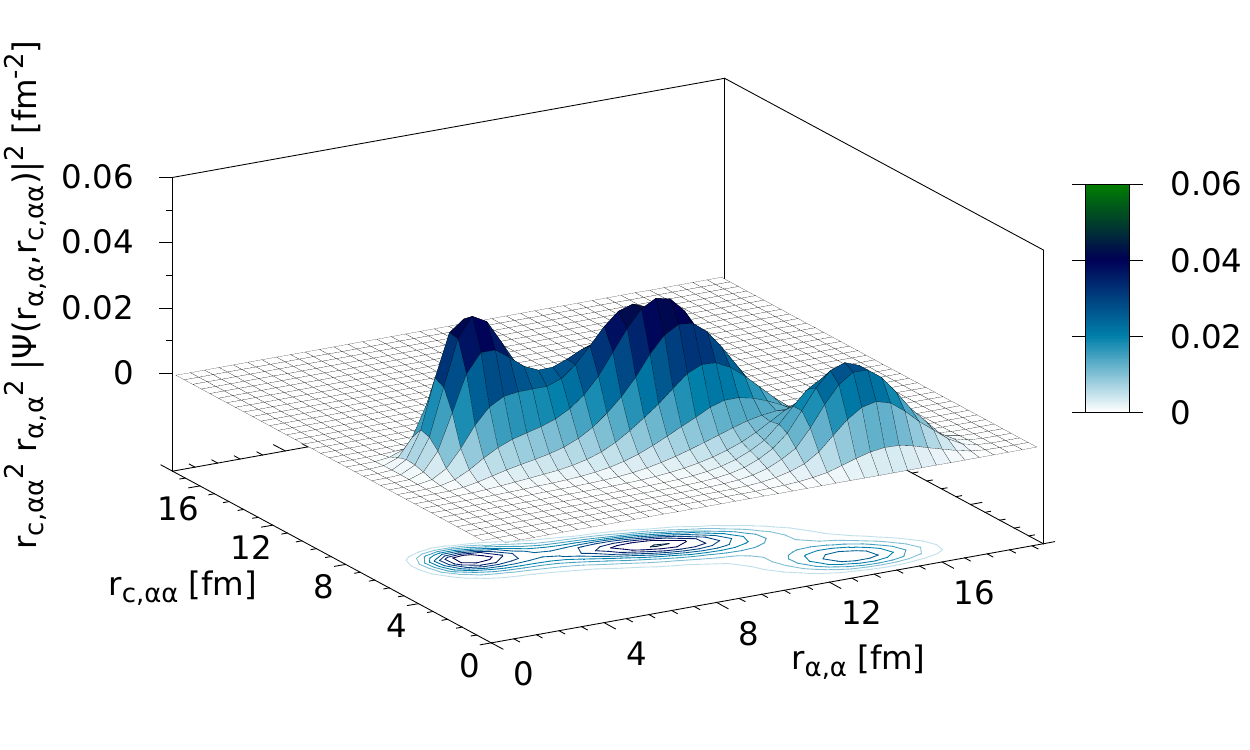}}

{\includegraphics[width=0.95\columnwidth]{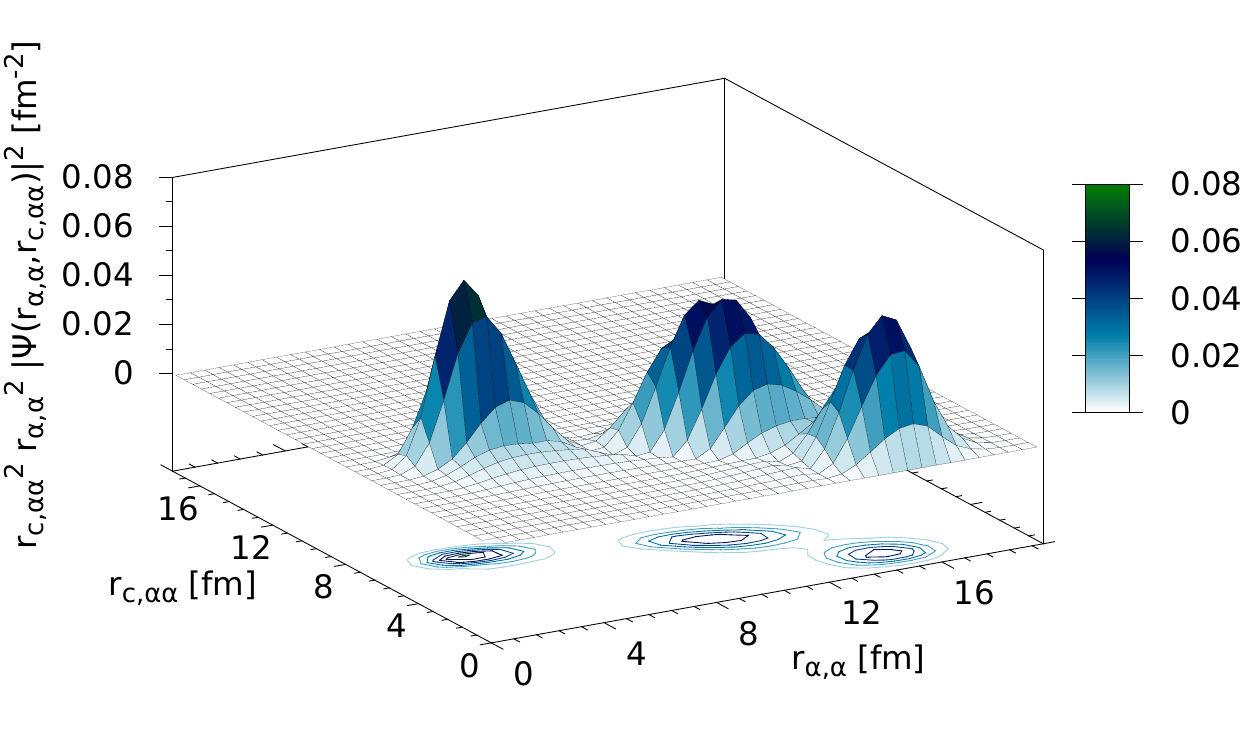}}

\caption{(Colour online) The same as top panel of Fig.~\ref{fig 3d 00}.
  for the second and third excited $2^+$ states.  \label{fig 3d 2}}
\end{figure}

The probability distributions for the lowest two $2^+$ states are very
similar to the $0^+$ distributions in Fig.~\ref{fig 3d 00}.  Also the
$\alpha$-core distributions are remarkably similar for the computed
higher-lying $2^+$ states.  The $\alpha$-$\alpha$ distribution for the
second excited $2^+$ state in Fig.~\ref{fig 3d 2} reveals a much broader
distribution. It is almost without peaks but with a ridge stretching
from $\alpha$-core distances between $10$~fm and $4$~fm with a
corresponding increase of the distance of the core from the
$\alpha$-$\alpha$ center of mass.  The third excited $2^+$ state is also
shown in Fig.~\ref{fig 3d 2} exhibiting distinct peaks at distances of
about $15$~fm, $11$~fm and $4$~fm.

\begin{figure}
\centering
{\includegraphics[width=0.95\columnwidth]{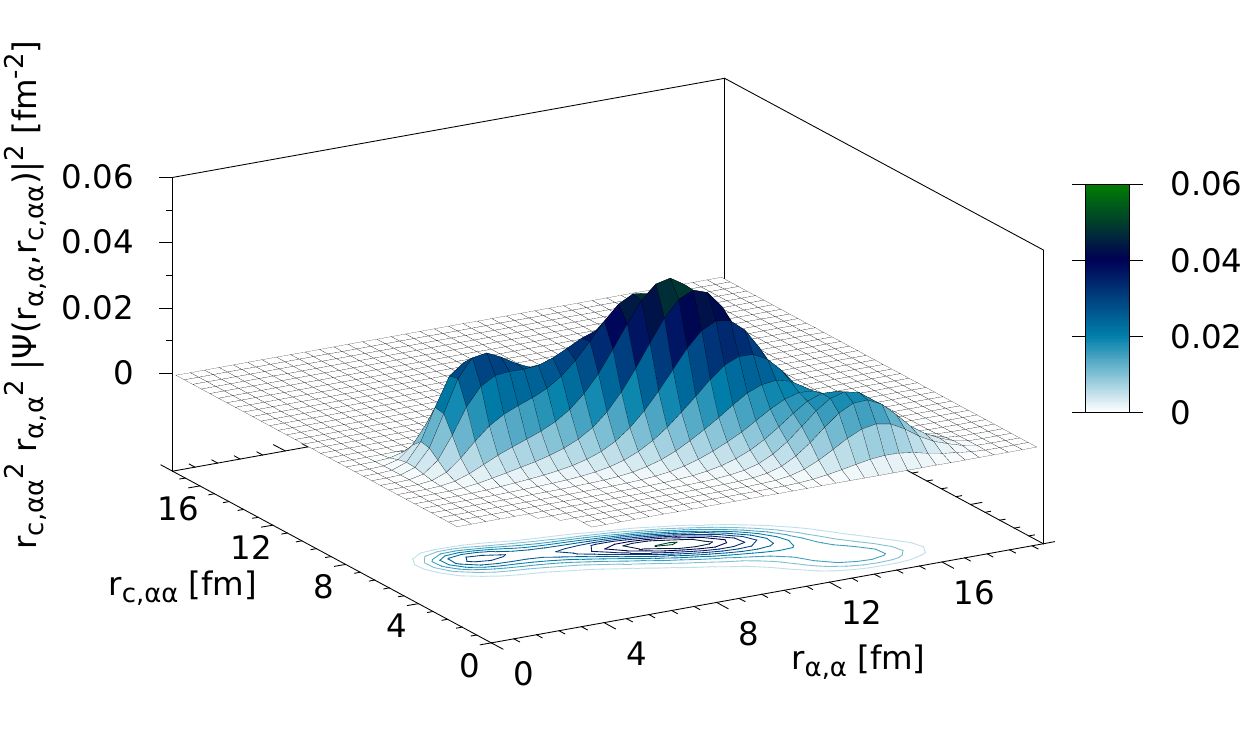}}

{\includegraphics[width=0.95\columnwidth]{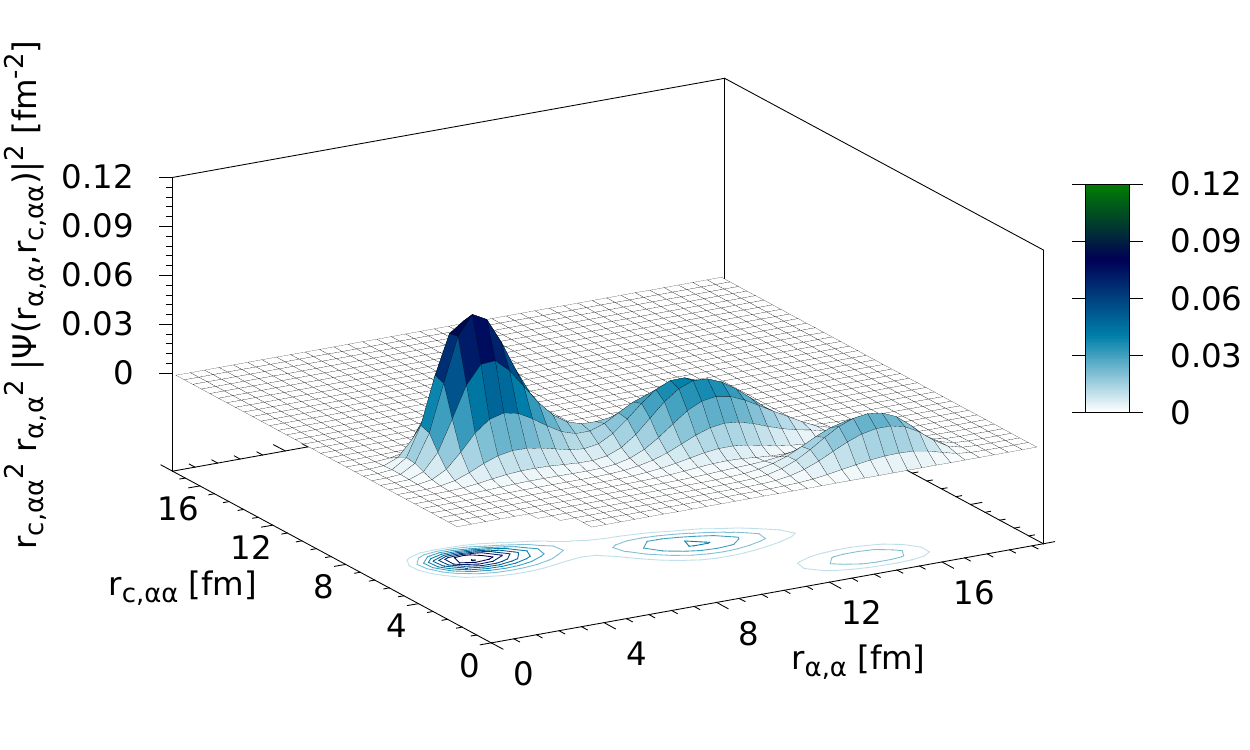}}

\caption{(Colour online) The same as top panel of Fig.~\ref{fig 3d 00}.
  for the second and third excited $4^+$ states.   \label{fig 3d 4}}
\end{figure}

The $\alpha$-core distributions for the $4^+$ states are again almost
indistinguishable from the previously computed distributions for the
other angular momenta, such as the example shown in the lower panel of
Fig.~\ref{fig 3d 00}.  The $\alpha$-$\alpha$ distribution for the first
excited $4^+$ state resembles the same distribution for both first
excited $0^+$ and $2^+$ states.  The distribution for the second
excited $4^+$ state is shown in Fig.~\ref{fig 3d 4}, where the peak
structure again is smeared out and the largest probability peak is at
around $11$~fm.  For the third excited $4^+$ state in Fig.~\ref{fig 3d
  4}, the $\alpha$-$\alpha$ distribution displays three peaks at
around $14$~fm, $10$~fm and $4$~fm.

In summary, the many different probability distributions all have the
remarkable property of one peak in the $\alpha$-core distance.  The
corresponding interaction is strong and essentially a surface
attraction due to centrifugal and Coulomb potentials.  In contrast the
$\alpha$-$\alpha$ distributions exhibit very large variations from one peak
to several peaks or rather smeared out distributions.  However, all
these distributions have fortunately a sharp cut-off at small
distances where the $\alpha$-particles would beginning to overlap.

The tempting interpretation in terms of simple geometric structures is
then only meaningful when one not too broad peak contains a large
fraction of the probability.  It may still be rewarding to look at
average distance properties as the root mean square radii given in
Table \ref{tab overview}.  Once more we emphasize that the $\alpha$-core
root mean square radius is remarkably constant for all states.  We can
then conclude that the $\alpha$-particles are located on spheres
corresponding to this radius around the core.

The average $\alpha$-$\alpha$ distances for the ground states of any angular
momentum are $4.8$~fm which is similar to, although about $1$~fm less
than, the same quantity, $5.95$~fm for the two-$\alpha$ structure
of $^{8}$Be.  Combined with the $70\%$ of $s$-waves in all these
ground states we conclude that these three-body states resemble a core
plus $^{8}$Be in its ground state.  The first excited states of all
three angular momenta are also very similar to each other but now with
more than $70\%$ of $\alpha$-$\alpha$ $2^+$ structure and with a much larger
distance of about $12$~fm.  This structure is far from any excited
state of $^{8}$Be, and these structures in fact resemble a linear
structure with the core in the middle.

The second and third excited states exhibit much more complicated
structures which cannot be collected into one simple
configuration. However, they can be described as containing three or
maybe even four components each with different configurations.  The
resulting probability distributions are more smeared out but both
$^{8}$Be like ground state structures, linear $\alpha$-core-$\alpha$
chain-configurations, and intermediate structures are present in each
state.

\section{Observable consequences \label{sec res}}

The general structures for systems with weak binding of one- and two
$\alpha$-particles are discussed in the previous section.  We shall here
first compare to measured properties of $^{148}\text{Nd}$, which was considered in Sec.~\ref{sec nd148} as a general representative of $\alpha$-Borromean structures in relatively heavy nuclei. In the second subsection we shall discuss the results
obtained from fine-tuning the interaction parameters to be appropriate
for the only known (apart from $^{12}$C) even-even Borromean two-$\alpha$
nucleus, $^{142}$Ba.

\subsection{Properties of $^{148}$Nd $(^{140}\text{Ba}+\alpha+\alpha)$}

The spectra in Table \ref{tab overview} for the lowest energies of the
$0^+$, $2^+$, and $4^+$ states present a rotational sequence,
$(0.0,0.4,1.4)$, with rigid body moment of inertia, ${\cal I}$
corresponding to $\hbar^2/{\cal I} \approx 0.14$~MeV.  This implies a
distance, $r_{c,\alpha\alpha}$, of about $6.19$~fm between the core and the
center of mass of two $\alpha$-particles, which is almost identical to the value
derived from Table \ref{tab overview}.  Furthermore, this is in
complete agreement with indistinguishable  geometric properties for
these three $0^+$, $2^+$ and $4^+$ states shown in Fig.~\ref{fig 3d
  00} for the $0^+$.

The conclusion is that these states form a rotational band. The
schematic rules for rotational B(E2)-transition probabilities are then
obeyed for a core plus a two-$\alpha$ structure rotating around their
common center of mass.  The absolute values of the electromagnetic
transition probabilities are proportional to the intrinsic electric
quadrupole moment, $Q_0$, of the same structure.  For a relatively
heavy core we have $Q_0 \approx 8e r_{c,\alpha\alpha}^2 \approx
307$~e~fm$^2$, where $4$~e is the charge of the combined two
$\alpha$-particles.  The single-particle value, $Q_{sp} \approx 75
$~e~fm$^2$, is about four (the charge) times smaller, provided the
same radius is used in both estimates.

Let us now compare these numerical average results to measured values
for $^{148}$Nd \cite{nic14}. First, the observed
excitation energies do not follow the simple
rotational model predictions. The energies of the $4^+$ state is $2.5$ times larger than the energies
of the $2^+$ states. If anything this is closer to the vibrational
model value of $2$ rather than $3.3$ valid for rotations.  The
vibrational picture does not match any better by combining the
second $0^+$ and $2^+$ states.

Transition probabilities contain more detailed information about
structures, but only rather uncertain data are available for these
nuclei. For $^{148}$Nd the available measurements of B(E2)
values are, $B(E2;0\rightarrow 2) = 1.37$~e$^2$~b$^2$, $B(E2;2\rightarrow 4) =
0.784$~e$^2$~b$^2$, and the quadrupole moment, $Q^{(2)} = -1.46$~e~b, for
the $2^+$ state.  Transforming these transition values into the down
going probabilities we get the ratio $0.78 \times 25 /(1.37 \times 9)
= 1.58$ which is comparable to $1.41$ from the rotational model but
also not too far from the vibrational value of $2$.  The quadrupole
moment is related to the intrinsic quadrupole moment by $Q_0 = -
Q^{(2)} 7/2 = 511$~e~fm$^2$ where our model value of $307$~e~fm$^2$ is
$1.7$ times smaller than measured.

Considering the same effective potentials were used for all partial waves, an agreement within a factor of two is better than what could have been expected. This is in spite of the fact that the model forms a rotational spectrum, while the data do not contain simple, strictly rotational or vibrational features. It is also worth noting that the simple model produced an energy spectrum where the energy of the $4^+$ state is $2.5$ times larger than the energy of the $2^+$ state, same as for $^{148}\text{Nd}$. This suggests that $^{148}\text{Nd}$, and nuclei similar to it, might well be described as two-$\alpha$ structures in their low-lying states.

The model with the same average parameters in the radial effective
potential is independent of angular momentum and known to be very
inaccurate for odd parity states.  This average model can only
marginally distinguish between odd and even parity states, since the
centrifugal barrier varies continuously with orbital angular momentum.
Only the Bose character of the $\alpha$-particles is able to give small
differences due to parity.  On the other hand, low-energy nuclear
spectra with only very few exceptions are dominated by the positive
parity states while negative parity states are located at higher
excitation energies.  In nuclear few-body models this feature is
accounted for by partial wave (angular momentum and parity) dependent
effective potentials.  A proper comparison to data therefore involves
detailed input and careful search for suitable nuclei where the
few-body structure is possible.

\subsection{Properties of $^{142}$Ba $(^{134}\text{Te} + \alpha + \alpha)$ \label{sec ba142}}

The most tempting nuclei to investigate are Borromean two-$\alpha$
systems.  Searching the available masses for candidates we find only
one known even-even nucleus, $^{142}$Ba ($\alpha + \alpha + ^{134}$Te), of
that structure.  The exception of $^{12}$C ($\alpha + \alpha +
\alpha$) is special since the core also consists of an $\alpha$-particle.
Nuclear few-body models must assume decoupling of intrinsic and
relative cluster degrees of freedom.  Therefore the intrinsic degrees
of freedom preferably should be difficult to excite either by
weak couplings or by unreachable high excitation energy.

\begin{table}
\centering
\caption{The Woods-Saxon depths, $V_0$, reproducing the resonance
  energies, $E_{res}$, for the different angular momentum and parities,
  $J^{\pi}$, in $^{138}$Xe $(^{134}\text{Te} + \alpha)$ \cite{son03}. There is no known $5^{-}$ state in the $^{138}\text{Xe}$ excitation spectrum, so the same potential depth as for $3^{-}$ was used. The energy of the $0^+$ state is determined by the $\alpha$ separation energy through Eq.~(\ref{eq ba142 sep}). All energies are in
  $\si{\mega\electronvolt}$.  \label{tab xe138}}
\begin{ruledtabular}
\begin{tabular}{c *{8}{c}}
$J^{\pi}$      & $0^{+}$  & $2^{+}$   &  $4^{+}$ &  $6^{+}$  &   $1^{-}$     &   $3^{-}$   &  $5^{-}$    &  Other   \\
\hline
$E_{res}$      & 0.138    & 0.727     &  1.211   &   1.419   &   2.004       &   2.153     &  -          &          \\
$V_0$          & 24.423   & 24.568    &  25.787  &   28.2    &   22.641      &   23.746    &  23.746     &  24.568  \\
\end{tabular}
\end{ruledtabular}
\end{table}

The present case has $^{134}$Te as the core where the lowest excited
state is a $2^+$ state at 1.279~MeV \cite{son04}. By adjusting the partial wave interactions the polarization is fully included in the adopted effective potential on the two-body level. If present, an $\alpha$-cluster structure should be seen as
resonances in $\alpha$-core scattering, that is as $^{138}$Xe states
\cite{son03}. The energy of the $0^+$ ground state is then determined by the $\alpha$ separation energy
\begin{align}
E_{res}(0^+) = -\left( B(^{138}\text{Xe})-B(^{134}\text{Te}) - B_{\alpha}\right). \label{eq ba142 sep}
\end{align}
This provides the depth of the radial potentials for
each angular momentum and natural parity.  We choose the same radial
Woods-Saxon shape with the same radius and diffuseness parameters as
used above.  We only adjust the depth to reproduce the measured resonance
energies in $^{138}$Xe ($\alpha + ^{134}$Te).  The resulting values
are given in Table \ref{tab xe138}.

\begin{table}
\centering
\caption{The same as Table \ref{tab overview} for $^{142}$Ba $(^{134}\text{Te} + \alpha + \alpha)$ using the
  potential depths specified in Table \ref{tab xe138} for the individual partial waves. The column
  labelled $E_{exp}$ is the measured value from \cite{joh11}, where the energy of the $0^+$ state is determined by the two $\alpha$ separation energy. Two first excited states, indicated by $(F)$, are also included. The last five columns specifies the individual weights of the five lowest potentials.  
\label{tab overview3}}
\begin{ruledtabular}
\begin{tabular}{ l *{5}{c} *{5}{c}} 
             &           &    &    &            &         & \multicolumn{5}{c}{Weights $(\%)$} \\
\cline{7-11}
$J^{\pi}$ & $E_{exp}$ & $E_{cal}$    & $E_{cal}-E_{exp}$   &  $\langle r_{\alpha \alpha}^2 \rangle^{\frac{1}{2}}$  &   $\langle r_{\alpha c}^2 \rangle^{\frac{1}{2}}$   & 1 & 2 & 3 & 4 & 5  \\
\colrule
$0^{+}$   & -0.16   &  0.11   &     0.27  &    7.8     &    7.0  &   99   &  1   &  0   &   0   &   0   \\
$2^{+}$   &  0.20   &  0.50   &     0.30  &    7.5     &    7.1  &   99   &  0   &  0   &   1   &   0   \\
$4^{+}$   &  0.68   &  0.98   &     0.30  &    7.4     &    7.1  &   96   &  2   &  1   &   0   &   0   \\
$1^{-}$   &  1.17   &  1.11   &    -0.06  &    4.2     &    7.1  &   98   &  1   &  0   &   0   &   0   \\
$3^{-}$   &  1.13   &  1.37   &     0.24  &    4.3     &    7.1  &   96   &  2   &  1   &   0   &   1   \\
$2^{+}(F)$&  1.26   &  1.05   &    -0.21  &    5.9     &    7.1  &   0    & 98   &  1   &   1   &   0   \\
$6^{+}$   &  1.31   &  0.90   &    -0.41  &    6.6     &    7.2  &   95   &  3   &  2   &   0   &   0   \\
$0^{+}(F)$&  1.38   &  1.59   &     0.22  &    9.3     &    7.1  &   2    & 96   &  1   &   0   &   0   \\
\end{tabular}
\end{ruledtabular}
\end{table}

The energies and sizes of the three-body eigenstates are given in Table
\ref{tab overview3}, together with the distribution of weights on the
different adiabatic potentials, and the experimentally measured energies \cite{joh11}. Here the ground state energy is determined by the two $\alpha$ separation energy. Also included in Table \ref{tab overview3} is the difference between the calculated and the measured energies.

The absolute values of the calculated energies are seen to be displaced by roughly $0.3 \, \si{\mega\electronvolt}$ for four of the five lowest states. A slight displacement is not surprising as no attempt has been made to account for explicit three body effects. A distinct three-body potential could be added, but it would be an ad hoc addition adjusted to fit the desired spectrum. More interesting are the relative distances between individual levels in the calculated spectrum, and they agree very well with relative distances between the experimental measurements. The calculated relative distances between the $0^+$, and the $2^+$, $4^+$, and $3^-$ states only differ by about $0.03 \, \si{\mega\electronvolt}$ from the relative distances in the experimental spectrum. The only exception among the lowest states is the $1^-$ state, which does not agree with the shifted spectrum. Its absolute value actually agrees more closely with the experimental value. This agreement is most likely coincidental, and merely the result of counteracting offsets. It may be of interest that in the symmetry classification of the corresponding states in the related system $^{12}\text{C}$ \cite{mar14} the lowest $1^-$ state appears in a different "band" than the $0^+$, $2^+$, $3^-$, and $4^+$ states.

For the three remaining, higher lying states, the deviations become more erratic. The difference between the calculated and the experimental values is no longer close to constant. This might indicate a limit to the model given by the core excitation energy.

The weights of the individual potentials are as expected from the results of the previous sections. Each state is dominated by a single adiabatic potential. The lowest potential dominates for all non-excited states, while the second potential dominates for the first excited states. The coupling between different potentials must then be very weak, even when the potentials of each partial wave are adjusted individually.

\begin{figure}
\centering
{\includegraphics[width=0.95\columnwidth]{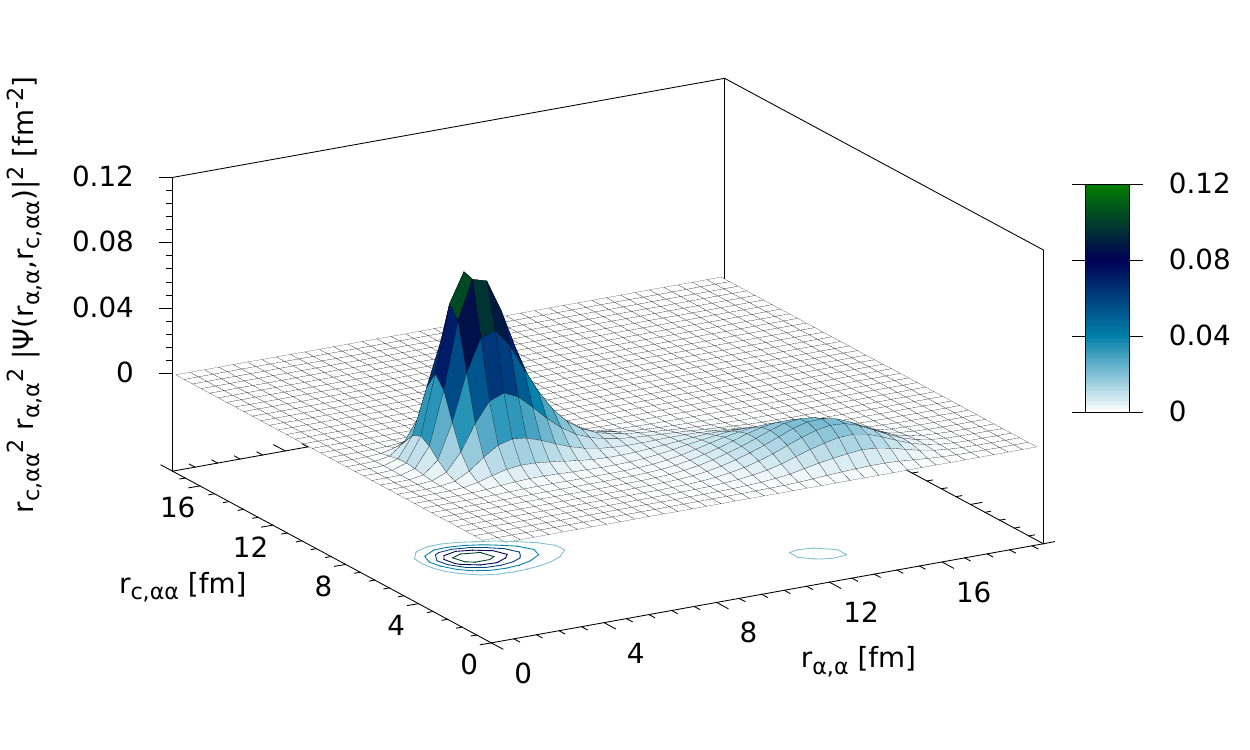}}
{\includegraphics[width=0.95\columnwidth]{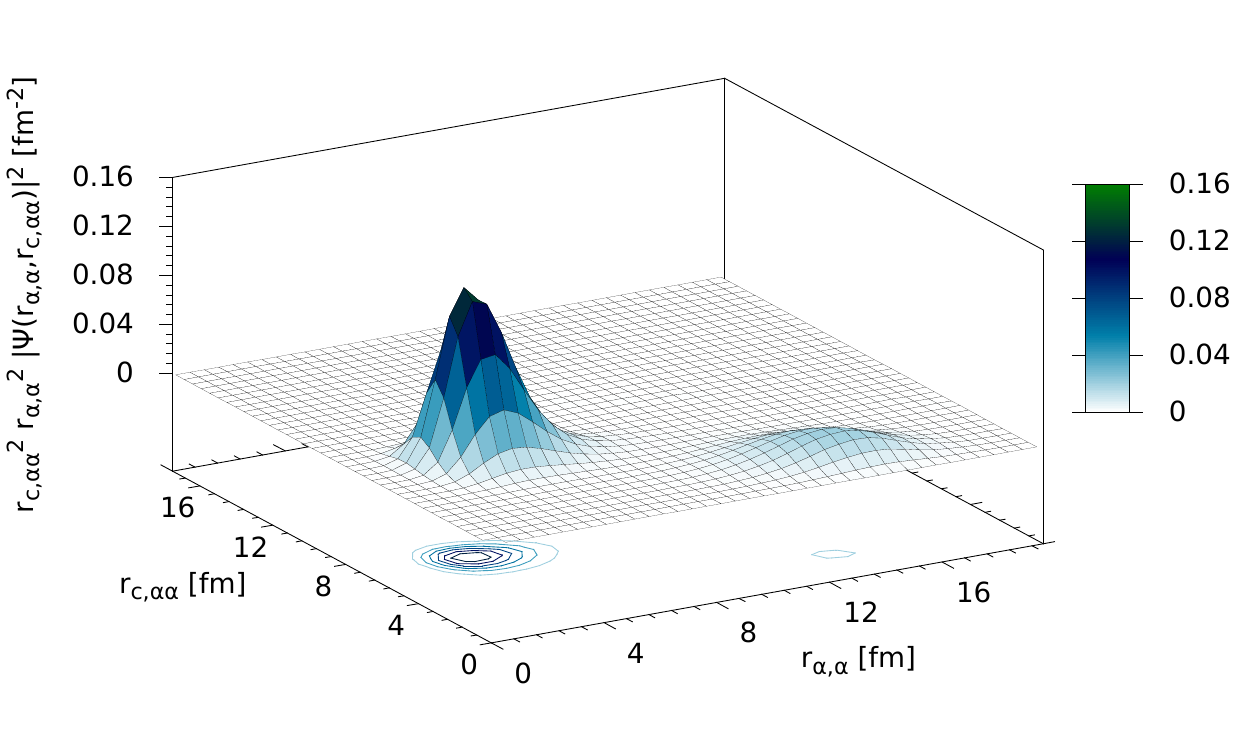}}
{\includegraphics[width=0.95\columnwidth]{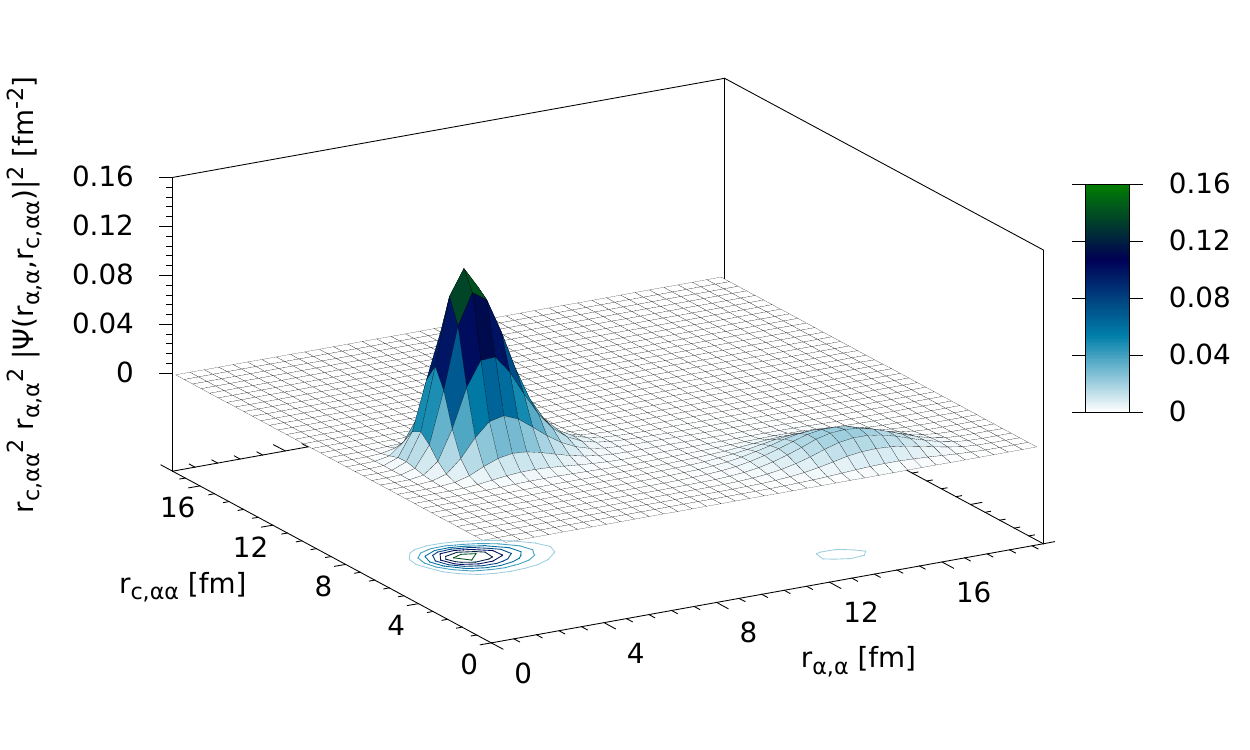}}
{\includegraphics[width=0.95\columnwidth]{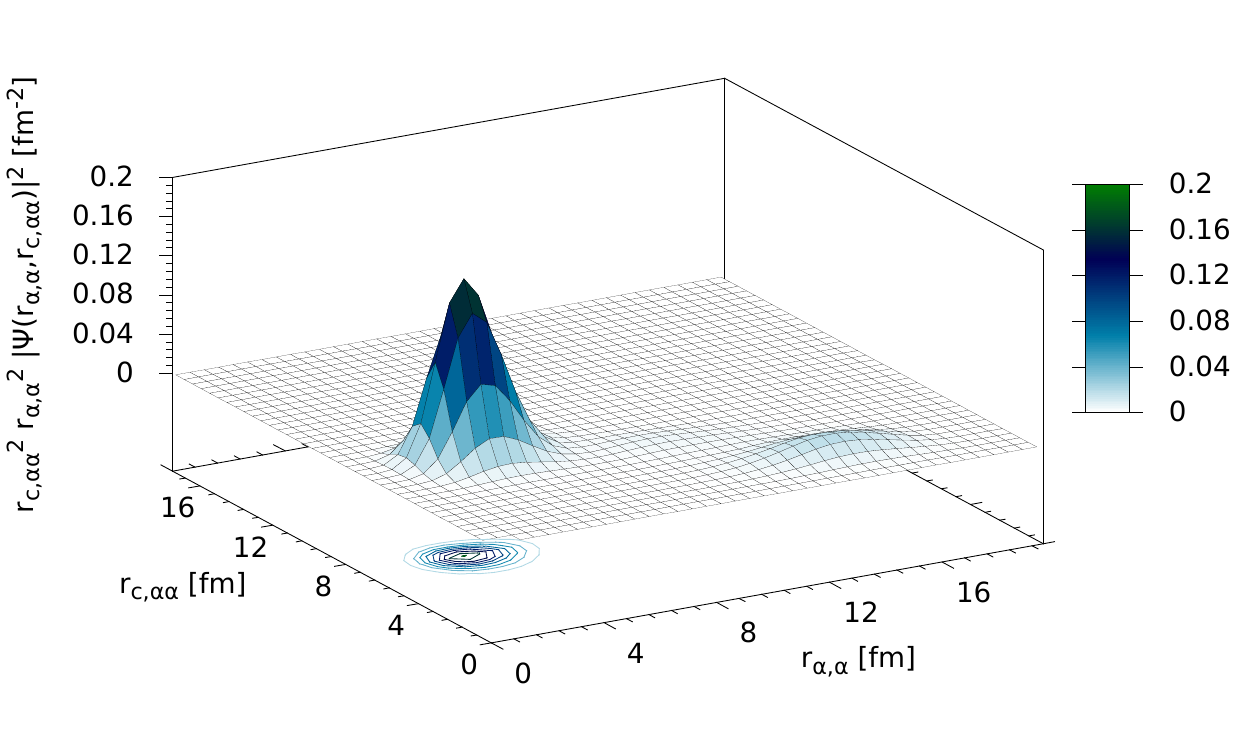}}
\caption{(Colour online) The same as top panel of Fig.~\ref{fig 3d 00}.
  for the lowest $0^+$, $2^+$, $4^+$, and $6^+$ $^{142}$Ba states.  
\label{fig ba+142}}
\end{figure}

The average $\alpha$-core distance is again constant at around $7.1 \, \si{\femto\metre}$, which implies that the $\alpha$ particles are still placed at the surface of a sphere around the core. The probability distributions between the core and the $\alpha$-particle are also identical to the distribution seen in the lower part of Fig.~\ref{fig 3d 00}, and is therefore not included.
The average $\alpha$-$\alpha$ distances, on the other hand, are very different from the ground state values in Table \ref{tab overview}, at least for the even parity states. However, the average values are somewhat misleading. In Fig.~\ref{fig ba+142} the probability distributions for the $0^+$, $2^+$, $4^+$, and $6^+$ states are shown. The same peak as in Fig.~\ref{fig 3d 00} at an $\alpha$-$\alpha$ distance of roughly $4 \, \si{\femto\metre}$ is seen for all three states. The large average values are caused by the appearance of a much smaller peak at an $\alpha$-$\alpha$ distance of about $13 \, \si{\femto\metre}$. This almost constitutes a line structure, with $\alpha$ particles on opposite sides of the core.

\begin{figure}
\centering 
{\includegraphics[width=0.95\columnwidth]{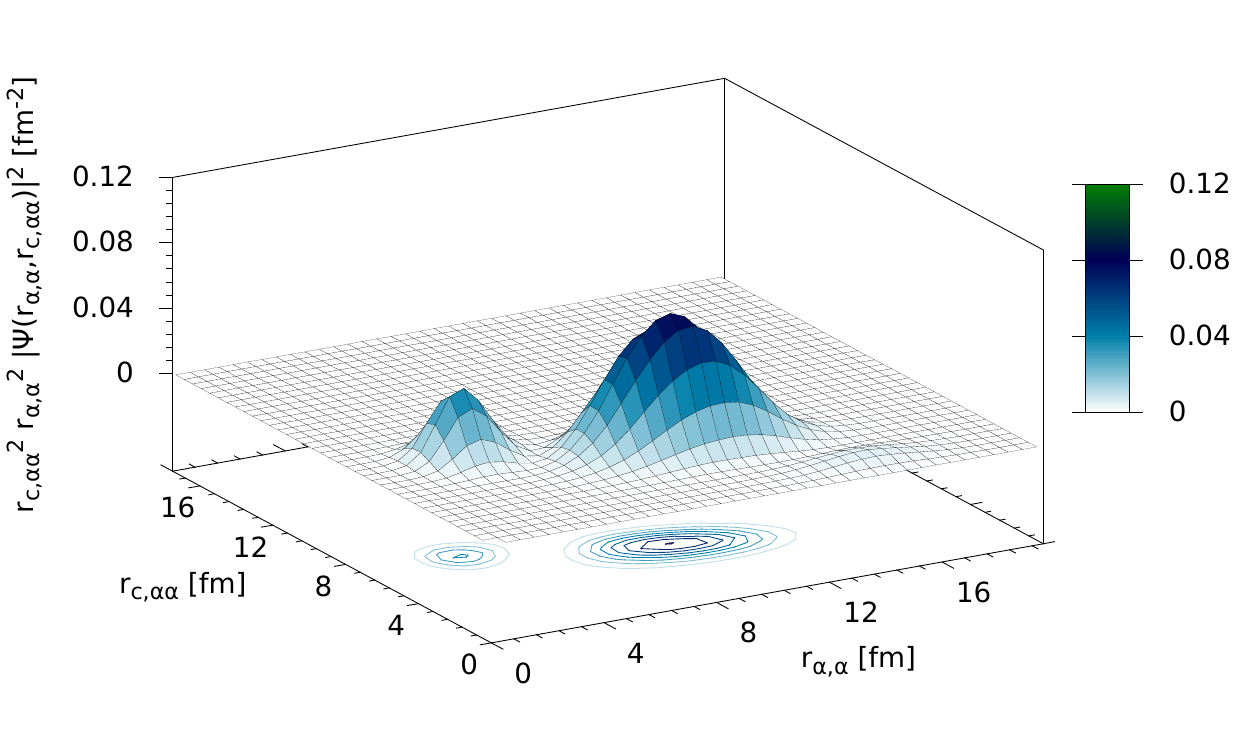}}
{\includegraphics[width=0.95\columnwidth]{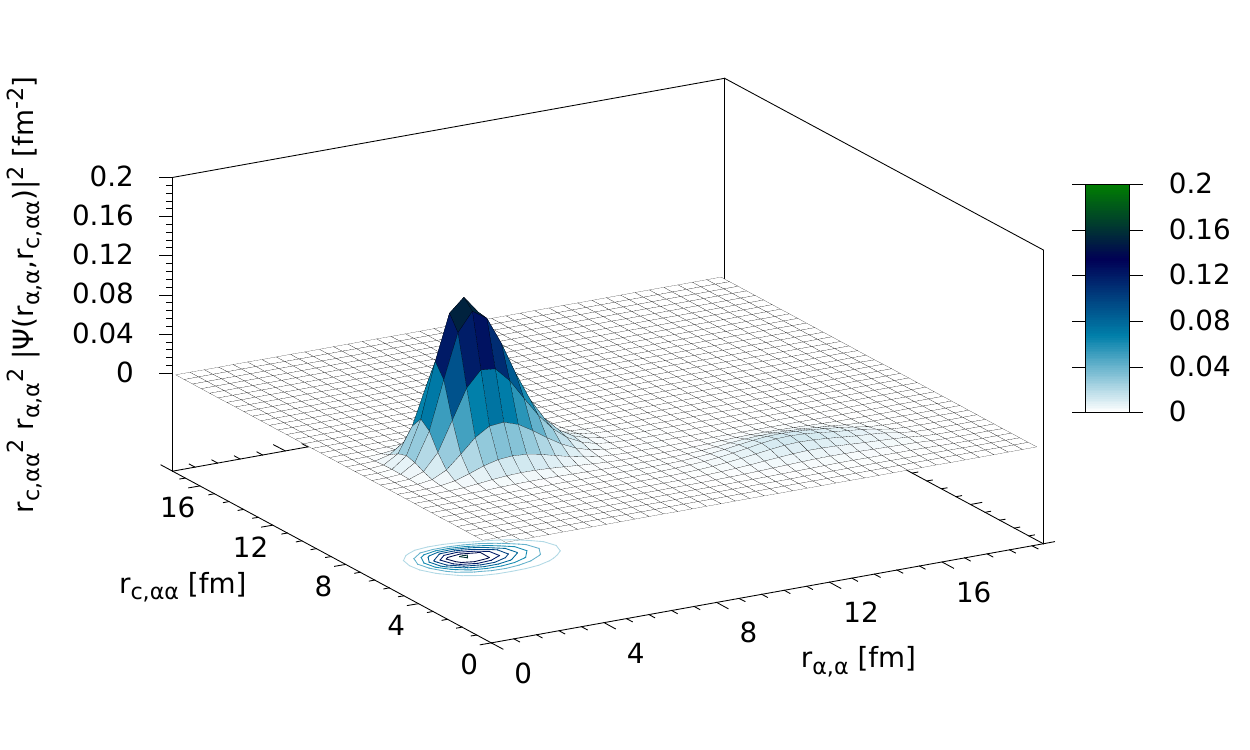}}
\caption{(Colour online) The same as top panel of Fig.~\ref{fig 3d 00}.
  for the first excited $0^+$ (above), and $2^+$ (below) states of $^{142}$Ba.  
  \label{fig ba+142 ex}}
\end{figure}

The probability distribution for first excited $0^+$ and $2^+$ states are seen in Fig.~\ref{fig ba+142 ex}. The distribution of the first excited $0^+$ is similar to the top panel of Fig.~\ref{fig 3d 30}, only with the large peak at the slightly smaller $\alpha$-$\alpha$ distance of $10 \, \si{\femto\metre}$. The distribution of the first excited $2^+$ state is unusual compared with the first excited states examined previously. It is almost identical to the distribution of the $2^+$ ground state seen in the second panel of Fig.~\ref{fig ba+142}.

\begin{figure}
\centering 
{\includegraphics[width=0.95\columnwidth]{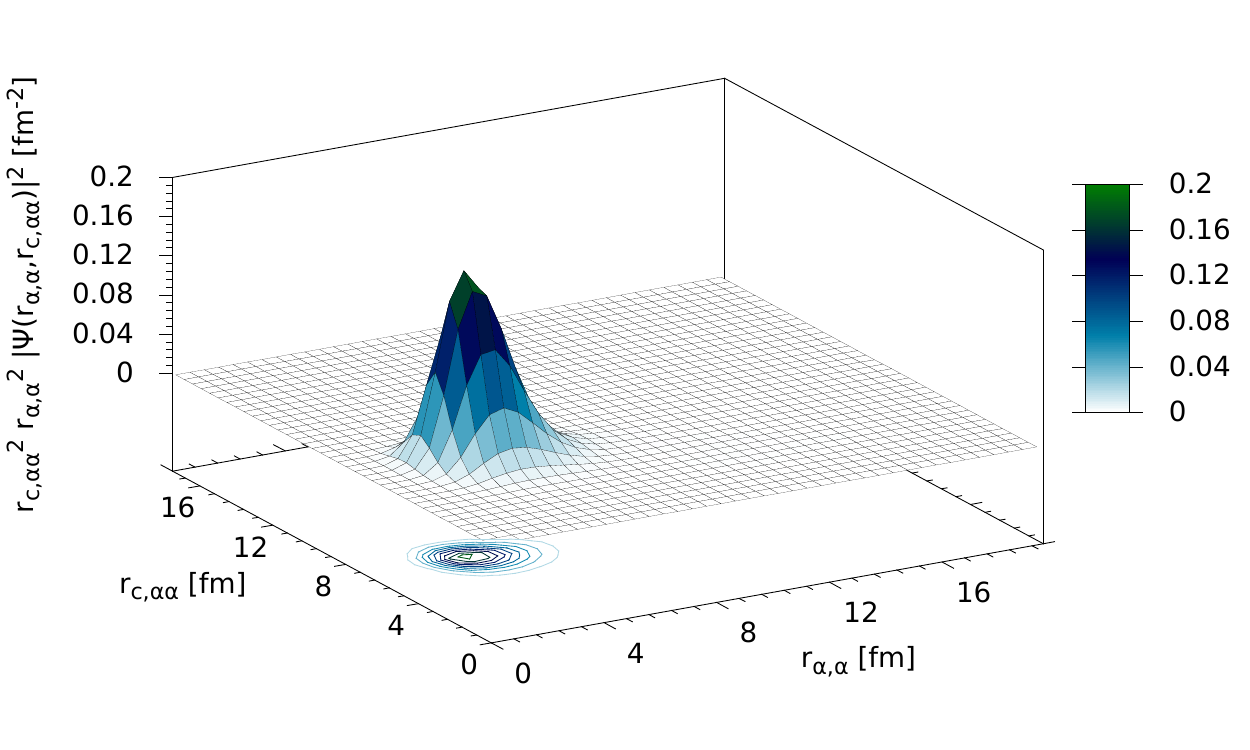}}
{\includegraphics[width=0.95\columnwidth]{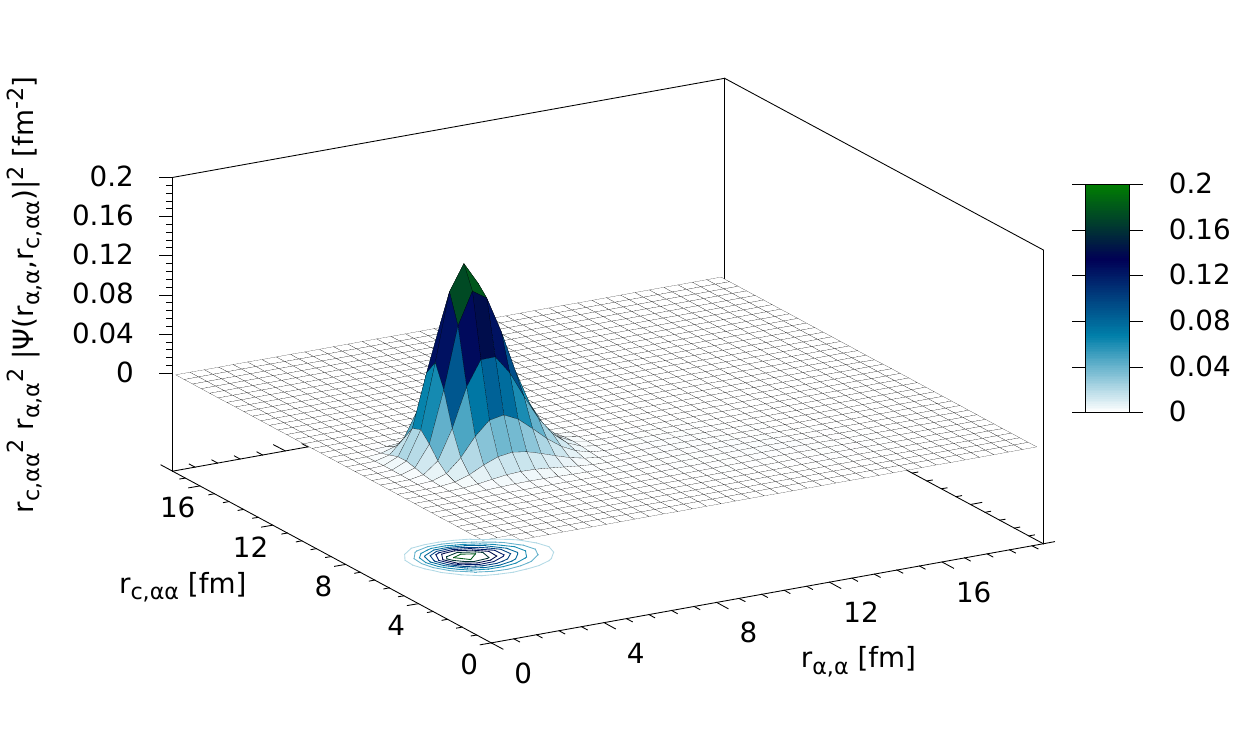}}
\caption{(Colour online) The same as top panel of Fig.~\ref{fig 3d 00}.
  for the lowest $1^-$, and $3^-$ states of $^{142}$Ba.  
  \label{fig ba-142}}
\end{figure}

On the other hand, the average distances for the odd parity states agree reasonably well with the earlier results. The probability distribution for the $1^-$, and $3^-$ states are seen in Fig.~\ref{fig ba-142}. They are identical to the upper part of Fig.~\ref{fig 3d 00}. The single peak is then well described by the average distance in Table \ref{tab overview3}.

The contributions from the different partial waves are listed in Table \ref{tab ba142}. The overall tendencies are the same as in Table \ref{tab 2p 4}, but the specific weights have changed slightly. The relative angular momentum between the two $\alpha$ particles is dominated by $s$-waves for all but the $6^+$ state, although to a lesser extent than before. Particularly interesting are the odd parity states which were not included earlier. The $1^-$ state is the only state which have a dominating contribution from $p$-waves to the relative $\alpha$-core angular momentum. This could be part of the explanation as to why the $1^-$ state deviates from the other low lying states. Likewise, the $3^-$ is the only state to have a significant $f$-wave contribution to the $\alpha$-core angular momentum. For the $3^-$ state there is a roughly even contribution from $s$-, $p$-, $d$-, and $f$-waves. The higher lying $6^+$ and $5^-$ states have a much more scattered distribution of partial waves, in particular for the $\alpha$-core system. However, these results are less reliable, as the states are outside the energy region, which can the model can reasonably be expected to cover.

\begin{table}
\centering
\caption{The same as Table \ref{tab 2p 4} with $^{142}$Ba for the states from Table \ref{tab overview3}
\label{tab ba142}}
\begin{ruledtabular}
\begin{tabular}{c| *6{c}}
$J^{\pi}$   &     Jacobi      & $l_x$ &   $l_y$      & $K_{max}$    &   G           &    F           \\
\colrule
$0^{+}$    & $\alpha$-$\alpha$   &   0   &   0          &    80     &  0.80         &   0.62         \\
           &                     &   2   &   2          &    60     &  0.18         &   0.30         \\
           &                     &   4   &   4          &    50     &  0.02         &   0.07         \\
\hline
$0^{+}$    & $\alpha$-c          &   0   &   0          &    100    &  0.70         &   0.27         \\
           &                     &   1   &   1          &    80     &  0.04         &   0.02         \\
           &                     &   2   &   2          &    60     &  0.17         &   0.67         \\
\hline
\hline
$2^{+}$    & $\alpha$-$\alpha$   &   0   &   2          &    70     &  0.62         &   0.06         \\
           &                     &   2   &   0          &    70     &  0.20         &   0.44         \\
           &                     &   2   &   2          &    50     &  0.11         &   0.46         \\
           &                     &   2   &   4          &    40     &  0.04         &   0.02         \\
\hline
$2^{+}$    & $\alpha$-c          &   0   &   2          &    70     &  0.35         &   0.00         \\
           &                     &   2   &   0          &    70     &  0.35         &   0.00         \\
           &                     &   2   &   2          &    50     &  0.09         &   0.17         \\
           &                     &   3   &   3          &    40     &  0.00         &   0.04         \\
           &                     &   2   &   4          &    40     &  0.02         &   0.05         \\
           &                     &   4   &   2          &    40     &  0.02         &   0.05         \\
           &                     &   4   &   4          &    30     &  0.01         &   0.30         \\
           &                     &   6   &   6          &    30     &  0.01         &   0.06         \\
\hline
\hline
$4^{+}$    & $\alpha$-$\alpha$   &   2   &   2          &    80     &  0.04         &                \\
           &                     &   0   &   4          &    50     &  0.63         &                \\
           &                     &   4   &   0          &    50     &  0.18         &                \\
           &                     &   2   &   4          &    48     &  0.08         &                \\
\hline
$4^{+}$    & $\alpha$-c          &   2   &   2          &    80     &  0.44         &                \\
           &                     &   4   &   0          &    50     &  0.17         &                \\
           &                     &   0   &   4          &    50     &  0.16         &                \\
\hline
\hline
$1^{-}$    & $\alpha$-$\alpha$   &   0   &   1          &    75     &  0.81         &                \\
           &                     &   2   &   1          &    65     &  0.09         &                \\
           &                     &   2   &   3          &    55     &  0.09         &                \\
\hline
$1^{-}$    & $\alpha$-c          &   0   &   1          &    75     &  0.22         &                \\
           &                     &   1   &   0          &    75     &  0.23         &                \\
           &                     &   2   &   1          &    55     &  0.15         &                \\
           &                     &   1   &   2          &    55     &  0.15         &                \\
           &                     &   2   &   3          &    55     &  0.05         &                \\
           &                     &   3   &   2          &    55     &  0.05         &                \\
\hline
\hline
$3^{-}$    & $\alpha$-$\alpha$   &   0   &   3          &    75     &  0.69         &                \\
           &                     &   2   &   3          &    55     &  0.26         &                \\
\hline
$3^{-}$    & $\alpha$-c          &   0   &   3          &    75     &  0.14         &                \\
           &                     &   3   &   0          &    75     &  0.15         &                \\
           &                     &   2   &   1          &    65     &  0.12         &                \\
           &                     &   1   &   2          &    65     &  0.12         &                \\
           &                     &   4   &   1          &    41     &  0.07         &                \\
           &                     &   1   &   4          &    41     &  0.06         &                \\   
\hline
\hline
$6^{+}$    & $\alpha$-$\alpha$   &   0   &   6          &   100     &  0.23         &                \\
           &                     &   6   &   0          &    90     &  0.06         &                \\
           &                     &   2   &   4          &    90     &  0.06         &                \\
           &                     &   2   &   6          &    90     &  0.52         &                \\
           &                     &   6   &   2          &    80     &  0.08         &                \\
\hline
$6^{+}$    & $\alpha$-c          &   0   &   6          &   100     &  0.14         &                \\
           &                     &   6   &   0          &    90     &  0.14         &                \\
           &                     &   2   &   4          &    90     &  0.08         &                \\ 
           &                     &   4   &   2          &    80     &  0.07         &                \\ 
           &                     &   4   &   6          &    40     &  0.04         &                \\ 
           &                     &   6   &   4          &    40     &  0.04         &                \\ 
           &                     &   6   &   6          &    50     &  0.14         &                \\                 
\end{tabular}
\end{ruledtabular}
\end{table}

The calculated dipole and quadrupole transition probabilities are presented in Table \ref{tab trans}. Included are both the results for the two-body $^{138}\text{Xe}$ $(^{134}\text{Te} + \alpha)$ and the three-body $^{142}\text{Ba}$ $(^{134}\text{Te}+2\alpha)$ systems. The transition probabilities are given by 
\begin{align}
B(E\lambda; J \rightarrow J^{\prime}) = Q_0^2 \langle J 0 \lambda 0 | J^{\prime} 0 \rangle^2,\label{eq trans}
\end{align}
where $\langle J 0 \lambda 0 | J^{\prime} 0 \rangle$ is the Clebsch-Gordan coefficient coupling the states $J$ and $J^{\prime}$. In the two-body system $Q_0$ for the quadrupole transition is 
\begin{align}
Q_0 = \sqrt{\frac{5}{16 \pi}} 2 r_{\alpha c}^2 \frac{z_{\alpha} e m_c^2 + z_c e m_{\alpha}^2}{(m_{\alpha} + m_c)^2},
\end{align}
while for the dipole transition $Q_0$ is 
\begin{align}
Q_0 = \sqrt{\frac{3}{4 \pi}} r_{\alpha c} \frac{z_{\alpha} e m_c - z_c e m_{\alpha}}{(m_{\alpha} + m_c)}.
\end{align}
Here $m_{\alpha}$ and $m_c$ are the masses, $z_{\alpha}$ and $z_c$ are the proton numbers, and $e$ is the elementary charge. The $\alpha$-core distance used is $7.1 \, \si{\femto\metre}$, as given in Table \ref{tab overview3}. There are two $\alpha$ particles in the three-body system, so twice the $\alpha$ mass, and twice the $\alpha$ charge is used. Also the distance is replaced by the distance between the core and the $\alpha$-$\alpha$ system, $r_{c,\alpha\alpha}$. The value used is not the average value from Table \ref{tab overview3}, but the peak value of $6.8 \, \si{\femto\metre}$ from Figs.~\ref{fig ba+142} and \ref{fig ba-142}. 

Unfortunately, only a few experimental values are available at the moment, as seen in Table \ref{tab trans}. Both the dipole and the quadrupole transition probabilities are relatively small, but for different reasons. The intrinsic particle degrees of freedom do not contribute to the rotational motion, because of the mass difference, so small values of the quadrupole transition probabilities are inherent in cluster rotations. Considering first the quadrupole transitions in $^{142}\text{Ba}$, the  model values are seen to be around $1.8$ times too large. It should be noted that the distance enters in the fourth power, so changing it slightly will have significant impact on the result. As this distance is dictated by Coulomb and centrifugal barriers, it does to some degree depend on the chosen parameters. This makes the agreement surprisingly good. The ratio between the states is $0.24/0.34 = 0.71$ in the model, which is very much comparable to the ratio of $0.77$ for the experimental values. Very few other calculations are available, but specialized models, such as the interacting boson model (IBM) \cite{sub11}, specifically designed to calculate transition probabilities, do exists. The few experimental $B(E2)$ values for $^{142}\text{Ba}$ are reproduced more accurately by IBM, but the model struggles with other, similar transitions for neighbouring nuclei. For $^{138}\text{Xe}$ only one transition probability is known experimentally. The three-particle model value for this transition is identical to the experimental value, although the experimental uncertainty is quite large. 

\begin{table}
\centering
\caption{The electric transition probability for the states from Table \ref{tab overview3}. The first column specifies the nuclei, the second column the transition in question, and the third column the calculated result based on Eq.~(\ref{eq trans}). The available experimental values for $^{142}\text{Ba}$ \cite{joh11} and $^{138}\text{Xe}$ \cite{son03} are presented in column four. A dash indicates the value was unavailable. The dipole transition probabilities are in units of $e^2 \, b$, and the quadrupole probabilities are in $e^2 \, b^2$. \label{tab trans}}
\begin{ruledtabular}
\begin{tabular}{l c d{5} d{6}} 
Nuclei             &      $B(E\lambda; J \rightarrow J^{\prime})$ &  \multicolumn{1}{c}{Model}  &  \multicolumn{1}{r}{Experiment}  \\
\colrule
$^{138}\text{Xe}$  &      $B(E2;2 \rightarrow 0)$                &       0.075                 &   0.076(20)  \\
                   &      $B(E2;4 \rightarrow 2)$                &       0.11                  &         -    \\
                   &      $B(E2;6 \rightarrow 4)$                &       0.12                  &         -    \\
                   &      $B(E2;3 \rightarrow 1)$                &       0.097                 &         -    \\
                   &      $B(E1;1 \rightarrow 0)$                &       0.0075                &         -    \\
                   &      $B(E1;1 \rightarrow 2)$                &       0.015                &         -    \\
                   &      $B(E1;3 \rightarrow 2)$                &       0.0097                &         -    \\
                   &      $B(E1;3 \rightarrow 4)$                &       0.013                 &         -    \\
$^{142}\text{Ba}$  &      $B(E2;2 \rightarrow 0)$                &       0.24                  &       0.145(4)   \\
                   &      $B(E2;4 \rightarrow 2)$                &       0.34                  &       0.188(12)   \\
                   &      $B(E2;6 \rightarrow 4)$                &       0.37                  &         -    \\
                   &      $B(E2;3 \rightarrow 1)$                &       0.30                  &         -    \\
                   &      $B(E1;1 \rightarrow 0)$                &       0.026                 &\multicolumn{1}{r}{$1.1(6) \cdot 10^{-6}$}\\
                   &      $B(E1;1 \rightarrow 2)$                &       0.052                 &\multicolumn{1}{r}{$2.0(10) \cdot 10^{-6}$}\\
                   &      $B(E1;3 \rightarrow 2)$                &       0.034                 &         -    \\
                   &      $B(E1;3 \rightarrow 4)$                &       0.045                 &         -    \\
\end{tabular}
\end{ruledtabular}
\end{table}

The absolute values of the dipole transitions are off by a factor $10^{-4}$, and does not resemble the experimental values. This is not surprising, as the small values are a result of the giant dipole resonances, which are not accounted for in this model. However, in spite of the large experimental uncertainties, the ratio between the transition probabilities is still a relevant test of the model, as the $1^{-}\rightarrow 0^+$ and $1^{-} \rightarrow 2^+$ transitions have almost equal branching ratios. The model's transition probabilities have the ratio $0.026/0.052 = 0.5$, which is very close to the experimental ratio of $0.55$. 

Another possible and very relevant test of the three-body model is to estimate the charge radius, and compare it with the measured value. The experimentally measured ground state charge radii are $\langle r^2_{ch}\rangle^{1/2} = 4.895(8) \, \si{\femto\metre}$, $\langle r^2_{ch-c}\rangle^{1/2} = 4.757(4) \, \si{\femto\metre}$, and $\langle r^2_{ch-\alpha}\rangle^{1/2} = 1.676(3) \, \si{\femto\metre}$ for $^{142}\text{Ba}$, $^{134}\text{Te}$, and the $\alpha$ particle respectively \cite{ang13}. The charge radius of the entire system can be calculated as $\langle r^2_{ch} \rangle = Z_T^{-1} \sum_{i=1}^{Z_T} \langle r^2_{i}\rangle$, where $Z_T$ is the total charge of the nucleus. For our three-body system this can be rewritten as
\begin{align}
\langle r^2_{ch} \rangle 
= \frac{Z_{c}}{Z_T} \left( \langle r^2_{c} \rangle + \langle r^2_{ch-c}\rangle \right)
+ 2 \frac{Z_{\alpha}}{Z_T} \left( \langle r^2_{ch-\alpha}\rangle + \langle r^2_{\alpha}\rangle\right), \label{eq ch rad}
\end{align}
where $Z_c$ and $Z_{\alpha}$ are the charges of the core and the $\alpha$ particle, and $\langle r^2_{c} \rangle$ and $\langle r^2_{\alpha}\rangle$ are the mean square radii of the core and the $\alpha$ particle respectively. These expectation values are calculated as in the three-body solution
\begin{align}
\langle r^2_c \rangle &= \left( \frac{2m_{\alpha}}{m_c+2m_{\alpha}} \right)^2 \langle r^2_{c,\alpha \alpha} \rangle,  \label{eq cm c} \\ 
\langle r^2_{\alpha} \rangle &= \left( \frac{m_c + m_{\alpha}}{m_c+2m_{\alpha}} \right)^2 \langle r^2_{\alpha, c \alpha} \rangle. \label{eq cm a}
\end{align}

Using Eqs.~(\ref{eq cm c}) and (\ref{eq cm a}) in Eq.~(\ref{eq ch rad}) the result is $\langle r^2_{ch}\rangle^{1/2} = 4.96 \, \si{\femto\metre}$, which is only $0.07 \, \si{\femto\metre}$ larger than the measured value. It should be noted that neither the charge distribution, nor the potential radius used in the three-body calculations have been adjusted to reproduce this charge radius. Such a close agreement is much better than what could have been expected. 

In summary, the low-energy spectrum of $^{142}\text{Ba}$ is reproduced by the present three-body model as well as by comparable light cluster  models. Both the measured and the calculated spectrum is neither rotational nor vibrational in character. However, the charge radius and the quadrupole transition probabilities are reproduced surprisingly well. The structure of states can be described as two $\alpha$ particles just outside the surface of the core, and located either just over $4 \, \si{\femto\metre}$ apart (possibly as a $^{8}\text{Be}$), or at opposite sides of the core in an almost linear chain.

\section{Summary and conclusion \label{sec con}}

We discuss the possibility of finding Borromean nuclear systems with
heavy constituents. Crudely speaking, two two-body systems each with
$Z^2/A > 17$ (squared charge over mass numbers) do not bind, that is
such pairs have negative binding energy.  They are then potential
candidates for constituents in a Borromean system.  However, this is
impossible as a third nucleus first would have to be similarly heavy
in order not to bind, and second its addition should produce a bound
three-body system.  Therefore it is hard to avoid light nucleons or
$\alpha$-particles but they can still be combined with one heavy
core-nucleus.

We sketch the driplines for nucleons and $\alpha$-particles, and
conclude that it is only possible to form a Borromean system with one
medium heavy nucleus by combining with two $\alpha$-particles, two
protons, or one proton and one $\alpha$-particle.  In the present
investigation we focus on two $\alpha$-particles and a medium heavy
core-nucleus.  An $\alpha$-$\alpha$ effective potential is chosen to
reproduce all low-energy scattering properties. The $\alpha$-core
effective potential is chosen in the same spirit to reproduce only the
weakest bound two-body states.  If the energy is zero this nucleus is
at the $\alpha$-dripline, and a positive binding energy could allow more
bound states where the weakest bound, or slightly unbound, is
appropriate as $\alpha$-cluster structure in an excited state.  These
states may be appropriate when the lower-lying $\alpha$-core states are
forbidden by the Pauli principle due to the same nucleonic
constituents in both core and $\alpha$-particle.

The core plus two-$\alpha$ calculations are carried out by use of the
hyperspherical adiabatic expansion method of the Faddeev equations.
The total angular momentum does not have to be zero and the
contributing individual partial waves can as well be finite.
Therefore we find bound state solutions for a few relatively small
angular momentum values.  The adiabatic potentials are all remarkably
similar with the same minimum and barrier positions.  They are
repulsively diverging at small distances, then steeply increasing from
the intermediate minimum towards larger distances, and finally they
decrease as Coulomb interactions at very large distance.  The
different adiabatic potentials are about $1.5$~MeV apart from
each other at the minimum, and their curvatures correspond to a
zero point energy of several MeV.

The $\alpha$-core potential is chosen to allow four three-body bound
states with energies varying from about $-5$~MeV up to almost
zero for each angular momentum.  Each bound state is dominated at the
level of more than $70\%$ by one potential term.  This means that the
angular structure of each bound state is directly related to one
adiabatic potential.  Still, their partial wave decompositions are
much more complicated, but with $s$-waves as clearly dominating
$\alpha$-$\alpha$ structures for the lowest bound state for all angular
momenta. The second lowest state is dominated by the reverse (with respect to $l_x$ and $l_y$) compared to the lowest state.
 The two highest-lying bound states in contrast contain large
fractions of $\alpha$-$\alpha$ $d$-waves.  This might indicate that the
$\alpha$-$\alpha$ system changes relative structure from ground to first
excited state of $^{8}$Be.  However, their distance is too large for
the attractive interaction to contribute, and this structure therefore
has to be attributed to angular momentum and parity conservation.

The spatial distributions of $\alpha$-particles around the core for the
different bound states are revealing.  The first striking result is
that the probability distributions as a function of $\alpha$-core distances
in all states are located in rather narrow distributions at distances
corresponding to $\alpha$-particles at the surface of the core.  This is
in contrast to the much more varying $\alpha$-$\alpha$ distance
distributions.  The lowest bound states for all angular momenta
show a spatial $\alpha$-$\alpha$ distribution similar to the $^{8}$Be
structure, but with a slightly smaller average distance.  However, the
higher-lying bound states clearly contain several configurations,
where the largest component often resembles a linear structure with
the core between the two $\alpha$ particles.  In these excited states an
$^{8}$Be-like structure is present but the other components are
usually dominating and the total probability distribution is much more
smeared out than for the two lowest bound states.

Measurable quantities like the energies should contain
information of $\alpha$-correlations. However, this is exceedingly
difficult to extract from the background of all other effects
contributing to the total energies.  We therefore focused on
$\alpha$-cluster structures resulting from strong $\alpha$-correlations.
These structures can be detected by scattering experiments where
$\alpha$-particle dripline nuclei are the most obvious targets. Large
cross sections for two-$\alpha$ removal can be expected as measured in
\cite{aki13}.  In more details, an $^{8}$Be structure should emerge
from the lowest of our three-body bound states, and two
non-interacting $\alpha$-particles can be expected from the three
higher-lying of the four computed three-body states. Measuring the transition matrix elements between different states will also constitute a test of the model.

To compare in more details with measured quantities we followed the
standard procedure in few-body nuclear physics. We focused on the
even-even Borromean two-$\alpha$ nucleus, $^{142}$Ba, where removal of
two $\alpha$-particles leave the fairly inert core nucleus, $^{134}$Te.
For each partial wave we construct effective potentials adjusted to
reproduce the $\alpha$-core, $^{138}$Xe, two-body resonances. With these
potentials we calculated three-body energies and found a very good
agreement with the lowest states in the known $^{142}$Ba spectrum, although the spectrum was shifted slightly most likely due to the fact that pure three body effects were not accounted for explicitly. The radial structure showed that the $\alpha$ particles were placed on the surface of the core. The relative angular momentum between the $\alpha$ particles was dominated by $s$-waves. In addition, both the electric quadrupole transition probabilities and the charge radius were reproduced rather well. Based on these finding we predict that the
corresponding structures of these low-lying states are two
$\alpha$-particles in a $^{8}$Be configuration rotating with different
angular momenta at the surface of a sphere around the $^{134}$Te-core.  This system is the most
promising for exhibiting $\alpha$ clusterization in the ground state.

In summary, we investigated the structures of two $\alpha$-particles
surrounding a heavy core-nucleus in a three-body model.  The
assumptions are that $\alpha$-clusters can be found with significant
probability in such nuclei.  We expect the most promising places in the
nuclear chart are at the $\alpha$ dripline where Borromean two-$\alpha$
structures are experimentally established by mass measurements.  These
nuclei should have relatively large sizes in their ground states when
the energies are close to zero.  This is the single most
important feature characterizing spatially extended halo structures,
which simultaneously enhance the possibility for decoupling of core
and $\alpha$-particle degrees of freedom.  These three-body structures
may also appear in excited states of nuclei where the $\alpha$-particle
is bound.  Then the $\alpha$ and core degrees of freedom may be mixed in
the ground states but decoupled in the excited state close to the
$\alpha$-threshold.

We have shown that Borromean two-$\alpha$ structures are possible at the
$\alpha$ dripline.  The $\alpha$-particles at the surface of the
core-nucleus would produce rotational spectra with the corresponding
simple transition probabilities.  Strongly enhanced $\alpha$-removal
cross sections would also be a signal.  One interesting perspective is
that similar proton-$\alpha$-core structures should be characteristic
features of ground states when proton and $\alpha$ driplines are close to
or intersect each other.

This work was funded by the Danish Council for Independent Research DFF Natural Science and the DFF Sapere Aude program.

\end{document}